\documentstyle[12pt, aaspp4] {article}
\def\etal{\it et al.~\rm}
\begin{document}
\title{Weak Lensing by High-Redshift Clusters of Galaxies - I: Cluster
Mass Reconstruction}
\author{D. Clowe\altaffilmark{1,2}, G.A. Luppino, N. Kaiser, and
I.M. Gioia\altaffilmark{3}}
\affil{Institute for Astronomy, University of Hawaii, 2680 Woodlawn Drive, 
Honolulu, HI 96822}
\altaffiltext{1}{Visiting Astronomer at the W. M. Keck Observatory, jointly 
operated by the California Institute of Technology and the University of 
California}
\altaffiltext{2}{Also Max-Planck-Institut f\"ur
Astrophysik, 85740 Garching, Germany}
\altaffiltext{3}{Also Istituto di Radioastronomia del CNR, Via Gobetti 101, 
40129 Bologna, Italy}

\begin{abstract}
We present the results of a weak lensing survey of six high-redshift
($z > 0.5$), X-ray selected clusters of galaxies.  We have obtained
ultra-deep $R$-band images of each cluster with the Keck Telescope,
and have measured a weak lensing signal from each cluster.  From the
background galaxy ellipticities we create two-dimensional maps of the 
surface mass density of
each cluster.  We find that the substructure seen in the mass
reconstructions typically agree well with substructure in both the
cluster galaxy 
distributions and X-ray images of the clusters.  We also measure the
one-dimensional radial profiles of the lensing signals and fit these
with both isothermal spheres and ``universal'' CDM profiles.  We find
that the more massive clusters are less compact and not as well fit by
isothermal spheres as the less massive clusters, possibly indicating
that they are still in the process of collapse.

\end{abstract}


\section{Introduction}

High-redshift clusters of galaxies are very powerful tools for testing
the predictions of cosmological and structure formation models.  The
mere existence of high-mass, X-ray luminous clusters at $z>0.5$
strongly constrains many models (Eke, Cole, \& Frenk \markcite{r13} 1996) while the
presence of a $\sim 10^{15} M_{\odot}$ cluster at $z\sim 0.8$ makes unlikely many
$\Omega_0 = 1, \Lambda = 0$ cold dark matter models (Luppino \&\ Gioia
\markcite{r30} 1995; Bahcall, Fan, \&\ Cen \markcite{r01} 1997; Henry
\markcite{r20} 1997).  Even stronger constraints can be placed on the
model once details of the clusters such as mass surface density,
cluster galaxy ages, amounts of sub-clustering, and mass-to-light
ratios are known (Crone, Evrard, \&
Richstone \markcite{r09} 1996; Trentham \& Mobasher \markcite{r42} 1998).

It has been only recently, however, that these high-redshift,
high-mass clusters have been able to be detected.  Optical surveys
tend to find low-mass clusters and often mistake the superpositioning
of unrelated groups of galaxies as clusters (Reblinsky \&\ Bartelmann
\markcite{r53} 1999).  Further, until recently,
the optical surveys were either too small to reasonably expect to find
a high-mass, high-redshift cluster or were not deep enough to detect the
high-redshift cluster galaxies.  

X-ray observations, however, have proven to be an efficient way to
detect these clusters.  The {\it Einstein} Extended Medium-Sensitivity
Survey (EMSS) found five $z>0.5$ high-mass ($>$ few $\times 10^{14} 
M_{\odot}$) clusters (Gioia \& Luppino \markcite{r18} 1994), and various {\it ROSAT}
surveys are finding many more (Henry \etal \markcite{22} 1997, Gioia \etal
\markcite{r19} 1999, 
Burke \markcite{r05} \etal 1998, Vikhlinin \etal \markcite{r44} 1998, Rosati \etal
\markcite{r37} 1998).  The
X-ray surveys not only provide clusters which are truly of high-mass,
and not a superposition of unrelated groups of galaxies, but also
provide a means to measure the masses of the clusters.  These masses
have been used to apply constraints to cosmological and structure
formation models (Henry \markcite{r20} 1997), but are subject to an uncertainty
in that the masses measured from the X-ray emission of the clusters
depend on the dynamical state of the cluster (Evrard, Metzler, \&\
Navarro \markcite{r49} 1996).

We, therefore, have undertaken an optical survey of these X-ray
selected, high-redshift clusters to perform weak lensing analysis on
the clusters.  Our primary goals in doing this survey are threefold:
First, we wish to measure the masses of the clusters using weak
lensing, which does not have a dependence on the dynamical state of
the cluster.  Second, we wish to determine the dynamical state of the
clusters and detect any substructure in the clusters.  Finally, we
wish to determine the redshifts of the faint blue galaxy (FBG)
population which is used in the weak lensing analysis.

We selected as our sample of clusters the five EMSS high-redshift
clusters (MS $0015.9+1609$ at $z=0.546$, MS $0451.6-0305$ at $z=0.550$, MS
$1054.4-0321$ at $z=0.826$, MS $1137.5+6625$ at $z=0.782$, and MS
$2053.7-0449$ at 
$z=0.583$), which were the only $z>0.5$ clusters published from a
serendipitous X-ray survey at the time, and one from the ROSAT North
Ecliptic Pole survey (RXJ $1716.6+6708$ at $z=0.809$) which was discovered
shortly after we began our survey (Henry \etal \markcite{r22} 1997, Gioia \etal
\markcite{r19} 1999).  We have already
published early results on the clusters MS $1137.5+6625$ and RXJ $1716.6+6708$
(Clowe \etal \markcite{r07} 1998).  In this paper we will present the
weak lensing data of the other four clusters as well as a correction
to the mass profiles published in Clowe \etal \markcite{r07} (1998).
We will discuss how the redshift distribution of the FBG population
can be constrained with the weak lensing observations in a future
paper (Paper II). 

In \S 2 we present the weak lensing theory used in our analysis.  Our
observations and details of the data reduction processes are given in
\S 3.  In \S 4 we discuss the properties of the clusters and the weak
lensing results.  Section 5 contains our conclusions.  Throughout this
paper, unless otherwise stated, we assume an $\Omega _0 = 1, \Lambda =
0$ cosmology, assume $H_0 = 100 h$ km/s/Mpc, and give all errors as 1
$\sigma$.

\section{Weak Lensing Theory}

It is well known that massive objects, such as clusters of galaxies, will
bend light-rays passing by them with their gravitational pull.  If there is
a good alignment between the positions of the background galaxy and the
center of mass of the cluster of galaxies then the background galaxy can
be strongly lensed.  In strong lensing, the galaxy is imaged into an arc,
or a series of arcs, along critical curves by magnifying the
galaxy's size tangentially to the cluster center of mass.  By measuring
the redshifts of the cluster and the arc, one can determine the mass
of the cluster interior to the arc, and also characteristics of the lensing
such as time delays between various images (Schneider, Ehlers, and
Falco \markcite{r38} 1992).

If the background galaxy is not well aligned with the cluster center of
mass lensing will still occur.  This lensing, however, will only slightly
distort the shape of the background galaxy, increasing the galaxy's size 
tangential to the cluster center of mass.  If the background galaxies were
all circular in shape then this weak distortion would be easily detected.
Because the background galaxy population has an intrinsic ellipticity
distribution (the dispersion of which is much higher than the change
introduced by the gravitational distortion), however, this weak lensing 
distortion can only be detected by looking for a statistical deviation from a 
zero average ellipticity tangential to a center of mass from a large
number of background galaxies.
Because of the intrinsic ellipticity distribution of the background galaxies
and that each image has a finite number density of background galaxies, 
weak lensing reconstruction is an inherently noisy process as the sample
of background galaxies will not have an average intrinsic ellipticity
that is precisely zero.  

Weak lensing, however, has two distinct advantages over other methods
traditionally used to measure cluster masses.  The first
advantage is that weak lensing provides a direct measure of the
surface density independent of the dynamical state of the cluster.
The other advantage is that the strength of the weak lensing signal
is directly proportional to the surface density and thus can, in
theory, be used to measure the surface density of structures much
smaller than, for instance, X-ray imaging whose emissivity scales as
the square of the density.  

\subsection{Basic Equations}

The goal of weak lensing observations is to measure the dimensionless
surface density of the clusters, $\kappa$, where
\begin{equation}
\kappa = {\Sigma \over \Sigma _{crit}}.
\end{equation}
$\Sigma $ is the two-dimensional surface density of the cluster, and 
$\Sigma _{crit}$ is a scaling factor:
\begin{equation}
\Sigma _{crit} = {c^2 \over 4 \pi  G}{D_s \over D_l D_{ls}}
\end{equation}
where $D_s$ is the angular distance to the source (background) galaxy, $D_l$
is the angular distance to the lens (cluster), and $D_{ls}$ is the angular
distance from the lens to the source galaxy.  The surface density
is related to the gravitational surface potential $\phi$:
\begin{equation}
\kappa = {1\over 2} \nabla ^2 \phi = {1\over 2} (\phi _{,11} + \phi _{,22}).
\end{equation}

The surface density $\kappa$ cannot, however, be measured directly from the
shapes of the background galaxies.  Instead, one can measure the mean
distortion in the galaxies by looking for a systematic deviation from
a zero average ellipticity.  From the distortion one can measure the
reduced shear $g$ (details given in the next section), which is
related to the shear $\gamma$ by
\begin{equation}
g = {\gamma \over 1-\kappa}
\end {equation}
(Miralde-Escude 1991).
In the weak lensing approximation, it is assumed that $\kappa << 1$.
Under that assumption and  that the background galaxies have an
isotropic intrinsic ellipticity distribution the distortion measured
in the galaxies can be translated to a shear by a direct scaling.
The shear $\gamma $ is related to the surface potential $\phi $ by:
\begin{eqnarray}
&\nonumber \gamma _1 = {1\over 2}(\phi _{,11} - \phi _{,22})&\\
&\gamma _2 = \phi _{,12}.&
\end{eqnarray}

While the lensing distorts, and enlarges, the background galaxy,
it preserves the surface brightness of the background galaxy.  As a
result, the total luminosity of the background galaxy is increased
proportional to the increase in its surface area.  The increase of the
size of each axis of the galaxy is simply $(1-\Lambda _i)^{-1}$ where
$\Lambda _i$ are the eigenvalues of $\phi _{,ij}$,
which is equivalent to $[(1-\kappa ) - \gamma ]^{-1}$ for the axis
tangential to the cluster center of mass, and $[(1-\kappa ) + \gamma ]^{-1}$
for the axis radial to the cluster center of mass for the weak lensing
regime.  Thus, the overall
increase in area, and the amplification of the background galaxy
luminosity, is:
\begin{equation}
A = |(1-\kappa )^2 - (\gamma ^2_1 + \gamma ^2_2)|^{-1}.
\end{equation}

One method to convert the measured shear field to a surface density is
to Fourier transform the data and
convert the Fourier transform of the shear,
$\tilde{\gamma }$, to that of the convergence, $\tilde{\kappa }$.  There
are a variety of ways to do this, but the most useful form is
\begin{equation}
\tilde{\kappa } = {(\hat{k}_1^2 - \hat{k}_2^2)\tilde{\gamma }_1 +
2\hat{k}_1\hat{k}_2\tilde{\gamma }_2\over \hat{k}_1^2 + \hat{k}_2^2}
\end{equation}
which has a flat noise power spectrum (Kaiser \markcite{r27} 1996).
One can then inverse  
Fourier transform to get an estimate of the surface density.  
This method, originally in Kaiser \&\ Squires \markcite{r25} (1993,
hereafter KS) and
hereafter called the KS93 algorithm, has a biasing at the edges
of the frame, particularly in the corners, as it assumes an infinite
spatial extent of the finite data field (Schneider \markcite{r60} 1995).  
Given that the images in our sample are centered on the clusters and
that structures of interest are typically restricted to the central
half of the images, this will not have a large effect on our results.
The KS93 algorithm, however, is only able to measure the convergence
to within an additive constant, the mean surface density of the field
(normally called the mass sheet degeneracy).
As the biasing at the edges would interfere with attempting to fit the
observed surface density with a chosen model to determine the mean
surface density of the field, we will use this method only to look for
substructure and not attempt to measure a mass.

One can also measure the surface density using aperture mass
densitometry, which measures the one dimensional radial mass
profile from an arbitrarily chosen center,  (Fahlman \etal
\markcite{r15} 1994).  The traditional statistic used is
\begin{equation}
\zeta (r_1,r_m) = \bar{\kappa }(<r_1) - \bar{\kappa}(r_1<r<r_{m}) = 
{2\over (1-r_1^2/r_{m}^2)}\int _{r_1}^{r_{m}} d\ln r \langle \gamma _T(r)
\rangle
\end{equation}
where $\langle \gamma _T(r) \rangle$ is the azimuthal average of the
galaxy ellipticity component measured with axes tangential and radial
to the center of mass at radius $r$ from the center of mass, and
provides a lower bound on the
surface density interior to radius $r_1$.  However, because this statistic
subtracts $\bar{\kappa }$ of the annulus outside $r_1$, the final measured
mass $M(<r_1) = \pi r_1^2\zeta (r_1)\Sigma _{crit}$ depends not only on the
strength of the detected lensing signal but also on the mass profile of the 
cluster.  In particular, any substantial substructure or secondary
core in the cluster would change from being included in the mass
estimate to being subtracted, at a reduced level, as $r_1$ moves
inward across the structure.  This would cause the measured slope of
the mass profile at this point to be shallower than what is actually
present.  One can easily modify the statistic to
\begin{equation}
\zeta _c(r_1,r_2,r_m) = \bar{\kappa }(<r_1) - \bar{\kappa }(r_2<r<r_{m}) =
2\int_{r_1}^{r_2}d\ln r \langle \gamma _T\rangle + {2\over (1-r_2^2/r_{m}^2)}
\int_{r_2}^{r_{m}}d\ln r \langle \gamma _T\rangle
\end{equation}
which subtracts a constant $\bar{\kappa }(r_2<r<r_{m})$ for a fixed inner
annular radius $r_2$ for all apertures in the measurement, thereby
removing any potential error from extended structures as mentioned above.
The measured mass profile $M(r_1) = \pi r_1^2\zeta _c(r_1)\Sigma_{crit}$
is now the mass of the cluster minus an unknown, but presumably small,
constant times $r_1^2$.  While the value of the additive constant
for aperture densitometry is not known, one knows what the constant physically
is, namely the average surface density of an annular region at some radius
around the center of the cluster. By modeling the mass distribution of the
cluster, one can hopefully get a good estimate of the value of this
constant.

All of the above equations assume that one can provide a direct
measure of the shear.  As one can only measure the reduced shear $g$
(eqn 4), all of these methods will measure a value for the convergence
which is too large.  Further, the ratio of the measured value to the
true value of the convergence increases with increasing convergence,
and thus a profile of the convergence would be measured to be much
steeper than it actually is.  The KS93 algorithm can be corrected for
this effect by first estimating the convergence field by assuming the
measured reduced shear is the true shear, and then iteratively
approximating the shear as $g(\vec{r})(1-\kappa (\vec{r}))$ where
$\kappa (\vec{r})$ is taken from the latest iteration (Seitz \&\
Schneider \markcite{r61} 1995).  While the aperture densitometry
profiles can also be corrected using an iterative technique, as one
uses a model to calculate the value of $\bar{\kappa }$ in the annular
region it is easier to simply calculate the $\zeta $ and
$\zeta _c$ for the models using the reduced shear for comparison to
the profile.

Even though $\zeta _c$ has a single constant subtracted
from all the bins, it is not necessarily always better to use than
$\zeta $.  A simple calculation of the signal to
noise ratios of the two statistics, assuming a constant background galaxy
density, only one source of lensing in or near the field, and a surface density
with a power law fall-off ($\kappa \propto r^{-n}$) gives
\begin{equation}
{\zeta _c/\sigma _{\zeta _c}\over \zeta /\sigma _{\zeta}} =
{(r_2^n-r_1^n)(r_m^2-r_1^2)\over r_2^n r_m^{2-n} (r_m^n-r_1^n)} +
{r_1^n (r_m^2-r_1^2)(r_m^n-r_2^n)\over r_2^n (r_m^n-r_1^n)(r_m^2-r_2^2)}
\end{equation}
where $r_m$, $r_2$, and $r_1$ are defined as in eqn 9.  As can be seen
in Figure 1, at large
values of $r_1$, the signal-to-noise ratio for $\zeta _c$ is worse than
that of $\zeta $, but as $r_1$ decreases it eventually becomes better than
that of $\zeta $.  The radius at which $\zeta _c$ has a better signal-to-noise
ratio depends on both the inner radius of the annulus, $r_2$, and $n$, the
power law fall-off of the surface density.  For an isothermal
sphere, $n=1$, $\zeta _c$ is generally worse in s/n ratio than $\zeta $ for all
but the innermost radii, but tends to do better for smaller values of
$n$.  

\subsection{$\kappa $ Reconstruction From Images}

The previous section deals mainly with how one can transform the gravitational
shear to a surface density.  For real data, however, one also needs a
mechanism to convert the background galaxies' ellipticities, which can
be measured directly from the images, to a gravitational shear value.
To measure the ellipticities of the galaxies we used a weighted second
moment of inertia such that
\begin{equation}
e = \left[ \begin{array}{c} 
	{Q_{11}-Q_{22}\over Q_{11}+Q_{22}}\\
	{2 Q_{12}\over Q_{11}+Q_{22}}
	\end{array} \right]
\end{equation}
where
\begin{equation}
Q_{ij} = \int d^2\theta W(\theta )\theta _i\theta _jf(\theta )
\end{equation}
where $\theta $ is measured relative to the center of the object (defined
by the centroid of the surface brightness), 
$f(\theta )$ is the surface brightness of the object, and
$W(\theta )$ is a weighting function (in this case a Gaussian)(KS).
One of the nice aspects of this method of defining the ellipticity is
that, in the weak lensing regime, at every point there 
will be some set of axes for which $e_1$ will be changed by the
gravitational shear while $e_2$ will remain the same as the pre-shear
value (or vice versa).  If one can determine the orientation of these
axes, such as the radial and tangential directions around a circularly
symmetric lens, then by averaging over many galaxies, one of the
ellipticity components can be used to 
measure the shear while the other would provide a measure of the
intrinsic ellipticity distribution of the galaxies.

It is, in theory, quite easy to convert the measured ellipticities
into a shear field.  Kaiser, Squires, \&  Broadhurst \markcite{r26}
(1995) (hereafter 
KSB) have shown that the shear field will change the ellipticity of
an object by
\begin{equation}
\delta e_{\alpha } = P_{\alpha \beta }^{\gamma } \gamma _{\beta}.
\end{equation}
The quasi-tensor $P_{\alpha \beta }^{\gamma }$ is an observable
quantity which can be calculated from the object image (KSB,
corrections in Hoekstra \etal \markcite{r23} 1998).  This calculation
was made in the 
weak lensing approximation ($\kappa << 1$); to use this in the inner
regions of massive clusters where the weak lensing approximation does
not hold, one needs to replace the shear $\gamma $ with the distortion
$g$.  Given that the mean ellipticity of the background galaxies
should be 0, one has
\begin{equation}
g_{\beta } = \left< {(P_{\alpha \beta }^{\gamma })^{-1} e_{\alpha}}
	\right>
\end{equation}
where the brackets indicate averaging in a chosen manner over the area
of interest.

\subsection{High Redshift Lenses}

Because $\Sigma _{crit}$ depends on $D_s/D_{ls}$, high redshift cluster lenses
have an additional complication over lower redshift lenses.  For low 
redshift ($z_l<0.3$) cluster lenses, $\Sigma _{crit}$ is effectively
the same for all background galaxies with redshifts several times
higher than the lens.  Thus, by taking an image deep enough to have
most of the background galaxies in the sample at $z\ga 0.7$, all of
the galaxies can be treated as being at the same 
redshift, and all will be magnified and distorted by the same amount.  
For higher redshift clusters, $z>0.5$ in particular, this is not the case, and
$\Sigma _{crit}$ continues to decrease fairly rapidly across the redshift
range in which most faint galaxies are thought to reside ($1<z<4$).
This variation in the lensing strength as a function of background
galaxy redshift provides not only some complications in the analysis,
but also a potentially useful tool to determine the redshift
distribution of the background galaxies, which will be discussed in paper II.

The first complication is that because $\Sigma _{crit}$ is no longer
constant for all the background galaxies one must compute a mean value
to convert the measured $\kappa$ to a surface density.  
From eqn 1, one has
\begin{equation}
\left< \kappa \right> = \Sigma \left< {1\over \Sigma{crit}} \right> =
\Sigma {{\int _{z_1}^{z_2}  n(z) dz\over \int _{z_1}^{z_2} 
{n(z)\over \Sigma _{crit}(z)} dz}}
\end{equation}
where $n(z)$ is the number of galaxies in the sample at redshift $z$.
Of course, eqn 15 assumes a knowledge of the redshift distribution of the
background galaxies, which is currently known spectroscopically to
only $z\approx 1$ (eg: Songaila \etal \markcite{r48} 1994, Koo \etal
\markcite{r50} 1996, Cohen \etal \markcite{r51} 2000).  Thus,
for clusters with $z>0.5$, for which background galaxies with $z>1$ must be
used to measure a weak lensing signal, one can only measure a mass for the
cluster by guessing at the redshift distribution of the background
galaxies or, if one has enough passbands, estimating the redshift of
each object using photometric redshifts (eg: Hogg \etal \markcite{r54}
1998).

The other complications created by the dependence of
the strength of the lensing on the redshifts of the background galaxies
are all due to the fact that, in addition to having the shapes of the
galaxies distorted, the sizes of the galaxies are enlarged while the surface
brightness remains constant, thereby magnifying the total luminosity of the
galaxies.  Because the greater the lensing strength the greater the
magnification of the background galaxy aperture luminosity, any constant
magnitude selection of background galaxy population will have the redshift 
distribution slightly altered from that of the unlensed background galaxy 
population (Broadhurst, Taylor, \&\ Peacock \markcite{r04} 1995). This
results in the lensed population having a larger fraction of
higher redshift galaxies, and thus the mean redshift of the background galaxy
population will have increased.  Because the strength of the lensing signal
depends on both the mass and the redshift of the cluster, both of these
will affect the $n(z)$ for the background galaxy population.  If no
correction is made for this effect, then the more massive clusters will be
measured to be even more massive than they truly are, and higher redshift
clusters will be measured to be more massive than the lower redshift clusters
of the same mass.

Another consequence of the magnification is that, because the strength of
the lensing signal increases as the distance from the center of the cluster
decreases, $\bar{\Sigma }_{crit}$ is not constant over the full extent
of a cluster (Fischer \&\ Tyson \markcite{r17} 1997).  
Without knowing the redshift distribution of the background galaxies
as a function of magnitude, one cannot even predict whether
$\bar{\Sigma }_{crit}$ will become larger or smaller as a function of
radius (Broadhurst, Taylor, \&\ Peacock \markcite{r04} 1995).  This effect,
however, should only be important near the cluster core (i.e. at large
$\kappa$).

All of these effects can be corrected for if the redshift distribution of the
background galaxies is known.  Currently, however, the redshift distribution
for field galaxies is known only to $z \approx 1$, which is much smaller than
the expected redshifts of the background galaxies in the images of this
sample.  Unlike Fischer \& Tyson \markcite{r17} (1997), we will not adopt a
theoretical model to predict the redshift distribution of the
background galaxy population.  Instead, we simply provide
our results with the caveats 
that the higher redshift clusters will have a higher measured surface
density than lower redshift clusters of the same mass and that there
might be a correction needed to the surface densities of the cluster
cores.  In both cases,
however, we expect that the corrections for galaxy magnification will
result in a change of less than $20\% $ of our measured surface densities.

\section{Observations}

Luppino and Kaiser \markcite{r32} (1997) (hereafter LK) show that for
high-redshift clusters,  
most of the weak lensing signal comes from the faint blue background galaxies.
The cluster galaxies, however, are quite red in color over the optical 
wavelengths.  Thus, in order to maximize the ability to distinguish cluster 
galaxies from the background galaxies used in the weak lensing analysis, the 
observations need to be taken in as red a band as possible, and in at least
two different bands to allow rejection of cluster and foreground galaxies
based on color.  Further, because LK did not see any significant lensing
until background galaxy magnitudes $I\geq 23.5$, the observations need at
least one color as deep as possible to get a signal-to-noise on the background
galaxies large enough to accurately measure the ellipticities of the galaxies.

The Low Resolution Imaging Spectrograph (LRIS, Oke \etal
\markcite{r36} 1995) at the 
Keck Observatory was judged to be ideal for this task.  It has a large
enough field of view ($6\arcmin \times 8\arcmin $) to allow imaging
outside the core of high-redshift clusters without having to mosaic
large offsets together, and has enough light gathering power (with the
10 meter primary mirrors of the Keck telescopes) to image the faint
background galaxies in only a few hours of integration per field.
However, because the $I$-band filter on LRIS was designed to not have
a long wavelength cutoff, the $I$-band images taken
with LRIS tend to exhibit large fringing effects.  As we were not sure this
fringing was stable and could be easily removed without affecting the quality
of the images, we choose to use the $R$-band as our primary observation band.
$I$-band and $B$-band images were obtained from the UH88$^{\prime
\prime }$ telescope using the 
Tek2048$^2$ CCD, which has a field of view and pixel size similar to that of 
LRIS.  

A list of the nights observed and data taken is given in Table
1.  All the data were taken using a ``shift-and-stare'' technique in
which short exposures (300s on Keck, 600s in $I$ and 900s in $B$ on
the UH88$^{\prime\prime}$) were obtained with small dithers ($\sim
15\arcsec$) between each exposure. Standard star observations and
calibration frames were also obtained for each night of observation.  
The data were obtained during nights which were mostly photometric.
The magnitudes of bright but unsaturated stars were used to determine
which images were non-photometric.  Those images which had stars 
significantly fainter ($>0.1$ magnitudes) than average were rejected
from the following analysis.   Typical seeing was $0\farcs6$ to
$0\farcs8$.

\subsection{Image Reduction}

The following image reduction routine was performed on all the data taken
after each observing run.  If all the nights had the same characteristics
(seeing, sky brightness, cloudiness, etc) then the entire run was processed
at the same time, otherwise each night was processed separately.  

The first step was the creation of a master bias frame from the bias 
frames taken at the beginning and ending of each night.  The pixel values of 
the master bias frame were the median of the pixels in each bias frame after
rejecting those pixels which were more than three $\sigma $ from the median.
This rejection removed any cosmic rays present in the bias frames before the 
median was computed.
The master bias frame was subtracted from each data and standard star
frame.

The over-scan region for each frame was averaged to a single column
and a linear least squares fit was performed on the column.  A three
$\sigma $ rejection routine was then performed on the over-scan column
to remove any values
increased by a star on the edge of the CCD, and the linear least squares fit
for the column was recomputed.  The fit was subtracted from each column
of the CCD data to remove any dark current and fluctuation in the bias value
over the course of the night.  In every case, the amount subtracted using
the over-scan regions after the master bias subtraction was less than 15 ADU,
or about 1$\% $ of the value of the bias.

The images were trimmed to the 2048$\times $2048 pixel physical CCD area,
and rotated, if necessary, so that north was increasing row numbers and
east was decreasing column numbers.  For the Keck images, which have the right
and left hand sides of the chip vignetted (at 0$^o$ rotation angle),
a clipping of the first 215 and last 233 columns was performed, with all the
pixels in those columns being set to a value which indicated that they should be
ignored for all future operations on the images.  The clipped region was 
slightly larger than that physically vignetted to remove any fringing
on the edges of the fields.

A flat field image was created for each band from the median of the
data images with sky level greater than one hundred times that of the read
noise of the CCD.  As with the creation of the master bias, while medianing
the data images, the pixels with values more then three times the standard
deviation of the median were rejected and the remaining pixel values were
re-medianed.  The flat field image was then divided by the mean value of
the image so that the average pixel value in the flat field image was unity.
Each data and standard star image was divided by the normalized flat
field for the band.  This removed quantum efficiency and through-put variations
across the image.

A bad pixel mask was created by combining the pixels which had a bias
value of greater than one third the full well capacity of the CCD with the
pixels which had a value in the flat field which was less than one fourth
the average value (or, in other words, masking out those pixels which would
give a signal-to-noise less than one half that of the average pixel on the 
CCD).  The bad pixel mask was then used to mask all of the data and standard
star images.  This masking set all of the ``bad'' pixels on the CCD to a value
which indicated to the averaging and analysis routines that the pixel should
be automatically excluded from any measurements.

The sky was then fitted and subtracted from all of the data and
standard star images.  This was accomplished by finding all the local minima
across the CCD and placing these in a smaller image, so that each pixel value
in the new image was the sum of the minima in the pixel range which was
binned to make the sky pixel.  A second image was also made which
was the number of minima used to create each sky pixel.  The two images were
then convolved with a Gaussian to remove high frequency noise and gaps in the
sky coverage.  The summed sky value image was then divided by the number of 
minima image to obtain a final value for each pixel, which was then subtracted
from the original image.
In addition to removing any low frequency sky variations,
this routine typically removed large wings of very bright, saturated stars
in the images as well as any large, low surface brightness objects which
might have been present.  It also would have removed any cD halos around
the brightest cluster galaxy of the clusters, but those regions were
edited  so that the sky around the BCG was an extrapolation
of the region outside that being edited (typically $\sim $20 pixel 
radius.)

A less robust version of the object finding routines described in
\S 3.3 was used to measure the centroid positions, fluxes, and half-light
radii for all the stars in the frames.  Data frames which showed a larger
half-light radius than the average (bad seeing) were excluded from the
sample.  The stellar centroid positions were used to calculate the offsets
and any distortion corrections.  For the UH88\arcsec \ $I$- and
$B$-band frames,
there was no significant distortion detected, and so the offsets were a
simple linear shift, with possibly a small rotation angle for data sets
taken on two different runs.  For the Keck $R$-band frames, however, a very
large distortion correction was needed to remove the effects of the curved
focal plane and the distortions introduced by the LRIS reimaging optics.  These
distortion corrections were calculated by mapping the stellar centroids onto
those for the summed $I$-band image, and fitting a bi-cubic polynomial
to minimize the differences in the two positions.  The stars selected
for the registration had $20<R<23$, which were bright enough to have
rms errors in the centroid less than 0.05 pixels ($\sim 0\farcs01$), 
determined from simulations of placing the stars on a Guassian random
noise background with the same rms as the observed sky noise and
calculating the centroid.  The internal rms error on the corrected positions
was typically $\sim 0.08$ pixels (compared to the positions of the
same stars in the other frames), and agreed with the positions in the
$I$-band image at $\sim 0.1$ pixels (which is the error in the
centroids in the $I$-band image due to higher sky noise).

The images were undistorted and shifted into the common registration 
frame as defined by the offsets calculated above.  Whenever a fractional
pixel shift was needed, a linear interpolation over the pixel and the
neighboring pixels was performed.  The amount added to each post-shift pixel
was calculated by integrating the interpolation over the region of the
pre-shift pixel overlapping each post-shift pixel.  The post-shift pixels
were then divided by the fractional area of the pre-shift pixels
contained within them in order to preserve surface brightness.  This technique
was tested to ensure that the second moments of the objects were not changed
by the fractional pixel shifts, except of course in the case of the distortion
correction.

The now registered images were then averaged after a 4$\sigma $
rejection routine was performed on all the stacked pixels to remove
cosmic rays
and moving objects.  It is interesting to note that
because the Keck telescopes have an alt-az mounting system, the diffraction
spikes seen
coming from bright stars because of deflection of the stellar light off
of the secondary mirror spider supports rotate as the field moves across the
sky.  As a result, the rejection routine removed the stellar spikes from the
$R$-band images, thus removing the chance that these might be detected as
objects by the detection routines described in \S 3.3.

The last step in the image reduction is to compare the fluxes in the
standard stars, all of which are taken from Landolt \markcite{r29}
(1992).  The exposure times for the standard stars were calculated so
that the charge in the central pixel would be
large, but still within the linear e$^-$/ADU regime on the CCD.  The stellar
flux was measured using an aperture large enough to include the radius at
which the flux becomes smaller than the sky noise, but small enough to not
include any nearby objects.  The sky level was calculated for each star by
creating a histogram of the pixel values in an annular region around the star
and fitting a Gaussian using a low weight for the pixel values greater than
the point at which the histogram drops below $\threequarters $ the peak value
on the high side of the median.  Any bad pixels within the half-light
radius of the standard star invalidated the star's use, while any bad pixels
outside the half-light radius but within the aperture radius for the star were
replaced with a linear interpolation from the surrounding pixels.  Finally,
for the Keck images, only those standards within the central 1024 pixels
were used to prevent a possible systematic error based on an incorrect
distortion correction.

\subsection{Object Detection and Analysis}

After creating an added image, the next step in the weak lensing analysis
is measuring the parameters of all the objects in the image.  We used the
IMCAT object detection and analysis package (available at
http://www.ifa.hawaii.edu/$\sim $kaiser/imcat) to perform most
of the steps described below.  

The first step was to detect all the various galaxies and stars in
the image.  This was done using a hierarchical peak-finding algorithm,
which smoothes
the image using progressively larger mexican-hat filters.  By comparing the
peak positions of the increasingly smoothed images, not only can the positions
of the objects be detected, but also a rough estimate of their size based on
the smoothing radius at which the peak is lost to the noise or combined with
another nearby peak (KSB).  
The technique is much better at finding small, faint objects and large, low 
surface brightness objects than the single aperture size scanning method
of FOCAS (Jarvis and Tyson \markcite{r24} 1981) and other similar programs, but tends to
produce a large number of detected noise spikes and often has problems with
both grouping together galaxies which are projected near each other and
breaking large spiral galaxies into smaller pieces.  The noise spikes, however,
are easily removed from the catalogs by size and signal-to-noise cuts,
and for purposes of this paper the large foreground spirals 
were excluded
from the analysis of background objects, so it does not matter in how
many pieces such galaxies are detected.  Further, because the detections of
a pair (or more) of nearby galaxies as a single object only occurs a small
fraction ($\approx 1\% $) of the time, these objects can also be excluded
from the catalogs without any biasing of the results or significant reduction
in the signal-to-noise of the weak lensing signal.

The next step was to determine the sky level around each object which 
was determined from an
annulus around the object with an inner radius of $16\times r_g$, where
$r_g$ is the smoothing radius at which the object achieved maximum
significance, and an outer radius of
$32\times r_g$.  This annulus was broken into four equal size regions by
radial divisions, and the mode pixel value for each quadrant was determined.
From these mode values the average sky level was determined, along with a
two-dimensional linear slope for the sky level.  A very large object,
such as a bright star or a foreground galaxy, could of course increase the mode
of the pixels in a quadrant, so usually all pixels within $3\times r_g$ of
another object were excluded from the mode calculation.

Once the sky value was known, an aperture flux and magnitude was
calculated.  The flux was
determined by summing all the pixels inside an aperture of $3\times r_g$
after subtracting the sky level as determined above.  This choice of 
aperture radius is large enough to count almost all of the light from the
object, but small enough to (usually) avoid including any light from nearby
objects.  Any pixels within $2.5\times r_g$ of another object were excluded
from the aperture.  Bad pixels were not corrected for but were noted in
the catalog of objects, and thus any objects with bad pixels in them could
easily be excluded (although this happened only for saturated stars, objects
overlapping a saturation spike of a very bright star, and objects on
the very edges of the summed image).  The aperture magnitude was then
calculated from the flux and a zero-point magnitude determined from the
standard stars.  A half-light radius, the radius at which the
integrated flux is $\onehalf $ the aperture flux, was also calculated.

The object shapes were determined by calculating the second moments of
the light distribution as given by eqn 12, using a circular Gaussian
with a standard deviation equal to $r_g$.  The centroid of the object
was calculated by minimizing the first moments of the weighted surface
brightness, using the same Gaussian weighting function as above.
The second moment is then
used to calculate the ellipticity of the object as given in \S 2.
The centroid position was then used as the position of the objects, instead
of that found in step 1.  This made little difference
in the weak lensing analysis as the two positions were different on average
by only 0.1 pixels.  Any objects with positional differences larger than
0.4 pixels ($\approx $ 4 times the error in the centroiding algorithm) were
excluded from the catalog as probably being either a multiple object detection
or an object having an unusual morphology which would not have a normal 
ellipticity.  This rejection tended to remove about $2\% $
of the objects in the catalog.

The next step in the analysis was to remove from the catalog of objects
noise spikes, saturated stars, and groups of objects detected as a single
object.  This was done by rejecting objects which had bad pixels inside
their aperture, objects which were extremely small (typically $r_g < 1.6$ 
pixels for 0\farcs 7 seeing and 0\farcs 22 pixels), objects which were overly 
large ($r_g > 10$ pixels), objects which were very faint ($R$-band magnitudes
$> 28$), and objects which had large ellipticities ($e > 0.5$).
The remaining objects in the catalog were then checked manually against the
summed image to remove any object groups which managed to pass all of the
above tests.

The final step in the object analysis was to obtain the sky level and
aperture photometry on the objects detected in the Keck $R$-band
images for the $B$- and $I$-band images.  

\subsection{Background Galaxy Selection}

Once a catalog of all the objects (stars, foreground, cluster, and background
galaxies) was generated, the next step was to isolate the background galaxies
and to correct for seeing and anisotropies in the point spread function.
The stars are very easy to separate from the galaxies by using a half-light
radius vs. magnitude plot.  As can be seen in Figure 3, the galaxies form
a broad diagonal swath in the plot while the stars are concentrated around
a single half-light radius value, and are clearly separated from the galaxies
for the brighter stars.  Thus by separating the objects in this ``finger'' on
the plot, the brighter stars can be isolated from the rest of the catalog.
Further, because any saturation of the star or the core of the star falling
into the non-linear response area of the CCD will increase the half-light
radius, it is easy to remove saturated stars from the star catalog.

The next step was to separate the background galaxies from the cluster and
foreground galaxies.
Because the redshifts of the galaxies in these images are only sparsely
sampled, it is impossible to fully distinguish between the three groups
of galaxies.  Figure 4 shows a color vs. magnitude plot for all the detected
objects in one of the fields, MS $1054.4-0321$.  At bright magnitudes, almost all 
of the detected galaxies (almost certainly foreground galaxies) lie in a 
narrow color band of $.6 < R-I < 1.1$.  Around $R = 22$ a second narrow
color band appears with $1.3 < R-I < 2.0$, which are the cluster galaxies (the
brightest cluster galaxy in this field has a color of $R-I = 1.6$).  The
$z\sim 0.55$ cluster galaxies are not quite as red as the $z\sim 0.8$ cluster
galaxies shown, with typical colors of $0.9 < R-I < 1.5$.  At
fainter magnitudes, starting around $R = 24$, the non-cluster galaxies begin
to break into two groups.  One of the groups stays in a narrow color range,
but the color as a whole gets redder, eventually merging with and then 
surpassing the $R-I$ colors of the cluster galaxies.  The other group forms
a broad swath of galaxies bluer than the foreground galaxies.  While it appears
from Figure 4 that the faint red galaxies outnumber the faint blue galaxies,
the opposite is in fact true.  Because the $I$-band images are not as deep as
the $R$-band images, most of the faint galaxies are not detected in the
$I$-band images but are assumed to be bluer than those which are
detected.

In order to have the background galaxies in each cluster analysis
drawn from the same redshift distribution, we applied the same
selection criteria for all of the fields.  In order to remove the
$z\sim 0.5$ cluster galaxies and have the background galaxy population
have most of its population at $z > 0.8$, we used those galaxies with
$23<R<26.3$ and $R-I<0.8$.  This resulted in a number density in the
background galaxy catalog between 33 and 42 galaxies/sq.~arcminute.
The spread in background number density is larger than the expected
variation due to Poissonian noise, and is due to incompleteness at the
faint end for the images with smaller exposure times and possibly
to the presence of faint blue cluster galaxies, particularly for the
$z\sim 0.8$ clusters, in
the final selection.  All further attempts at refining the selection using
additional colors resulted in a lowering of the signal-to-noise of the
lensing signal.

\subsection{Correction for Seeing and Distortion}

Once the background galaxy sample has been selected, the last step before
the weak lensing analysis programs could be applied was to correct the
ellipticities of the galaxies for atmospheric seeing and telescope distortion.
The primary effect of the seeing is that the ellipticities of the galaxies 
have been reduced because the original shape of the galaxy has been smeared
by the nearly-circular point spread function.  Small anisotropies in the
seeing along with aberrations (such as coma and wind shake) in the telescope 
focal plane can cause an apparent shear which must be removed
from the data, otherwise a false mass signal will be generated during the
weak lensing analysis.  

Because the stars are near point-sources before the light enters the 
atmosphere and they are not being lensed by the cluster in the
image, the shapes of the stellar profiles in the image are caused by the
effects that need to be removed from the galaxy profiles.  Thus, one can
measure the ellipticities and sizes of the stellar profiles and can use
these to correct the galaxy profiles.  
This can be done in a manner very similar to that used to describe how
galaxies would respond to an applied shear given in \S 2.2.  One can
calculate from each object a quasi-tensor $P_{\alpha \beta }^{s}$
such that
\begin{equation}
\delta e_{\alpha } = P_{\alpha \beta }^{s} p_{\beta }
\end{equation}
(KSB, corrections in Hoekstra \etal \markcite{r23} 1998).
Thus, $p_{\beta }$, which is an analog of the shear field $\gamma $
for the anisotropic smearing,
can, in theory, be derived from the ellipticities of the 
stars in the image.  The ellipticities of the galaxies near those stars could
be corrected to what they would be for a circular psf.
In practice, the shot noise
of the stars creates some noise in the second moments used to calculate the
ellipticities.  Thus instead of using each star to correct the galaxies
around it, we fit the ellipticities of the stars as a two-dimensional
polynomial as a function of position.  Each galaxy's position can then
be used to calculate the ellipticity of the stellar field at that spot,
and thus the correction to the galaxy's ellipticity could be
calculated.  Shown in Figure 5 are the ellipticities of the stars in
the MS $2053.7-0449$ field both before and after the fitted ellipticity correction.
The faint galaxies also have a large error in $P_{\alpha \beta }^s$ caused
by sky and shot noise on the galaxies, but simulations have shown that
a better recovery of the circularized psf ellipticity is obtained using each
galaxy's $P_{\alpha \beta }^s$ than by trying to calculate an ensemble average
for galaxies with a similar size, ellipticity, and orientation.  This is due
mainly to the fact that one can construct two objects to have the same size,
ellipticity (as measured by second moments), orientation, and total luminosity,
but have radically different morphology, and thus they would deform differently
under an applied smearing kernel.

Once the object shapes have been corrected to a circular psf, the next step
is to remove the dilution of the ellipticity caused by the smearing of the
object by the psf.  Because the $P_{\alpha \beta }^{\gamma }$ is measured for
the objects after the seeing has been applied, one cannot use the method of
\S 2.2 to obtain the shear.  LK have shown, however, that effects of
seeing can be removed using
\begin{equation}
g_{\beta} = \left< e_{\alpha} (P^{\gamma}_{\alpha \beta} - 
P^s_{\alpha \delta} P^{\gamma \ast }_{\delta \epsilon} 
{P^{s \ast }_{\epsilon \beta}}^{-1})^{-1} \right>
\end{equation}
where an asterisk denotes the value for the stars in the frame. 

Because the object
ellipticities have been altered since the $P$ values were calculated, the
measured $P$ values are no longer valid.
Calculating the correct $P$ values for the
ellipticity corrected objects is non-trivial (and near impossible for faint
objects because of the sky noise), so instead we use
${P^{\gamma}_{\alpha \beta}}^{\prime} = {1\over 2} (P_{11}^{\gamma } +
P_{22}^{\gamma }) \delta_{\alpha \beta}$ and ${P^s_{\alpha
\beta}}^{\prime} = {1\over 2} (P_{11}^s + P_{22}^s) \delta_{\alpha
\beta}$.  In doing this conversion, we are making two assumptions 
about how the P values change.  The first is that the off-diagonal terms are
small compared to the on-diagonal terms, and thus can be ignored.  The second
assumption is that the size of the objects does not change when doing the
ellipticity corrections, so the trace of the $P$ values remains the same after
the corrections.

This technique was tested on simulated data which was first sheared, then
convolved with a Gaussian psf.  The standard data analysis package was used
on both the sheared, pre-seeing image and the post-seeing image.  As can
be seen in Figure 6, the recovered shear from both images is nearly
identical.

\subsection{Simulations}

One of the goals of the study was to determine the dynamical state and
detect any substructure in the clusters.  The aperture densitometry
profiles will be somewhat useful in this regard as they can determine
if the radial profiles of the clusters are similar to those expected
from collapsed objects.  Of more use, however, would be to detect and
measure the mass of any structures not part of the cluster cores.  To
do this, however, the noise levels in the mass reconstructions must be
calculated.

The largest non-systematic noise source in weak lensing analysis is the
intrinsic ellipticity distribution of the background galaxy population.
Thus, the level of noise in the mass reconstructions depends primarily
on the background galaxy density.  It does, however, also have a
significant dependence on the image quality (how sharp the psf is).
Because of the inherent shot noise in the flux detected from the galaxies
and the brightness of the night sky, there is an unavoidable error in measuring
the second moments of the background galaxies, and thus their ellipticities.
Further, due to the fact that worse ``seeing'' (a larger psf) results in a
larger correction factor as described in \S 3.4, the error in corrected
ellipticities is greater for poorer quality images.  This increased error not
only reduces the measured signal, but also can create large noise spikes
from only a handful of galaxies in which the noise has tangentially aligned
their ellipticities about some point.  The shapes of these noise spikes
are usually small and round, although extended structures have been seen in 
some simulations.  They are most easily detected using aperture densitometry,
in which they show a signal only over a limited range of radii.  By using
simulations, however, we can compare the size and strength of features
seen in the cluster mass reconstructions to those in simulations and
get an estimate for the significance of the features.

An example of these simulations is shown in Figure 7.  These
simulations were created using 1960 galaxies (40 galaxies per
sq. arcminute) randomly placed on a 7\arcmin$\times$7\arcmin\ field.
Each galaxy was given an ellipticity drawn randomly from the pool of
all of the background galaxies in the six fields in the survey.  The
position and ellipticity of each galaxy were then altered to simulate
being lensed by a 1000 km/s singular isothermal sphere located in the
center of the field assuming $z_{lens} = 0.8$ and $z_{bg} = 1.5$.
Thus, all structures in these simulations which are not part of the
singular isothermal sphere lens are a result of noise.  The sizes of
these structures can then be compared with those in the cluster mass
reconstructions to get a measure of the significance of the structures.

\section{Cluster Properties}

In Table 2 we list the redshifts, X-ray luminosities (converted to a
rest frame 0.3 - 3.5 keV band) and temperatures, and the velocity
dispersion of galaxies for the clusters in the sample, all of which
are taken from the literature.  We also give the $R$ magnitude of the 
brightest cluster galaxy (BCG) within a 10$h^{-1}$ kpc radius circular
aperture and cluster galaxy number counts and Abell richness class.
All of the galaxies we chose as BCGs have been
spectroscopically identified as being a member of the cluster (Carlberg
\markcite{r06} \etal 1996; Donahue \etal \markcite{r11} 1998,
\markcite{r61} 1999; Gioia \&\ Luppino \markcite{r18} 1994).
The cluster galaxy number counts and corresponding Abell richness class
were measured following the
prescription of Bahcall \markcite{r02} (1981) in which all galaxies
within two magnitudes fainter than the third brightest cluster member
and less than 250$h^{-1}$ kpc from the BCG were counted.  The number
of background galaxies was estimated using the same magnitude
selection of galaxies at the edges of each image and scaled by the
ratio of areas between the two samples.  This estimate
was subtracted from the observed number counts around the BCG.  The
errors for the number counts in Table 2 are based on Poissonian noise
for the background galaxy counts and errors in the photometry of the
third brightest cluster galaxy and the fainter galaxies which could
cause galaxies to move into or out of the magnitude limits.  The third
brightest cluster galaxy was chosen by finding the third brightest
object within 250$h^{-1}$ kpc of the BCG which was extended compared to
the psf and had a $R-I$ color within .3 magnitudes of the BCG ($\sim
1.3$ for the $z\sim 0.55$ clusters and $\sim 1.6$ for the $z\sim 0.8$
clusters).

The KS93 mass reconstructions for the clusters are shown in Figures 8
($z\sim 0.55$ clusters) and 9 ($z\sim 0.8$ clusters).  The
reconstructions are 352\arcsec\ in width (1300$h^{-1}$ kpc at $z=0.55$
and 1450$h^{-1}$ kpc at $z=0.8$) and have been smoothed by a Gaussian
filter with a $17\farcs6$ standard deviation.  The number density of
background galaxies used to make these mass maps for each cluster are
given in Table 2.  Overlayed in contour on the mass maps
are X-ray images of the fields from ROSAT HRI observations, smoothed
by the same Gaussian filter as the mass map (Neumann \& B\"ohringer
\markcite{r35} 1997; Donahue \markcite{r10} 1996; Donahue \etal
\markcite{r11} 1998, \markcite{r61} 1999; Gioia \etal
\markcite{r19} 1999).  Also shown in Figures 8 and 9 are the
distribution of galaxies with R-I colors within .3 magnitudes of the
BCG weighted by $R$-band luminosity and by number.  Both images are
smoothed by the same Gaussian filter as the massmap.

A profile, centered on the BCG, of the reduced shear and aperture
densitometry for each cluster is shown in Figure 10.  The outer radius
of the profiles is the distance from the BCG to the nearest edge of
the R-band image.  The minimum radius from the BCG was chosen to be
100$h^{-1}$ kpc, which is large enough to still have a usable number
of background galaxies for shear estimation but is sufficiently far
from the Einstein radius (the radius at which $\bar{\kappa} = 1$ and
strong lensing occurs) that the approximation for the reduced shear
used in \S $2.2$ is still valid.  The reduced shear profiles were fit with both
``universal'' CDM profiles (Navarro, Frenk, and White \markcite{r34}
1996, hereafter NFW), integrated to a surface density (Bartelmann
\markcite{r03} 1996) and isothermal sphere profiles using a $\chi ^2$
determination for quality of fit and assuming that the background
galaxies lie in a sheet at $z_{bg} = 1.5$.  The parameters of the best
fit for each model as well as the $\chi^2$ and significance from a
zero mass model for each cluster are given in Table 2.
It should be noted that the NFW models are not robust fits,
as the two parameters can be adjusted against each other to some extent and
not severely decrease the quality of the fit, as can be seen in Figure
11.  While the parameters
describing the best fit depend on the assumed redshift of the
background galaxies, changing this assumed redshift will not
alter the quality of the fit of the profiles or the
significance from a zero mass model.  Changing the assumed redshift of
the background galaxies will also change the mass computed from the
measured $\bar{\kappa }$ in the aperture densitometry profiles. 

\subsection{MS $0015.9+1609$}

MS $0015.9+1609$ is the most well-known of the clusters in our sample.  It
was originally discovered by R. Kron in 1975 (Koo \markcite{r28} 1981)
and has served as 
a high-redshift cluster in most every survey since (eg: Dressler \etal
\markcite{r12} 1997, Smail \etal \markcite{r39} 1997, Yee \etal
\markcite{r47} 1996).  It was included in the EMSS,
which was a serendipitous and not targeted survey, because it was in the
field of another targeted cluster (Gioia and
Luppino \markcite{r18} 1994, hereafter GL).  A composite three-color
image of the cluster is given in Figure 12.

Optically, the cluster is very easy to recognize with three bright galaxies
in a line running north-east to south-west, the BCG being the central
galaxy of the three, surrounded by a large number of
fainter galaxies.  MS$0015.9+1609$ has the highest number of color
selected cluster galaxies in the sample, but these counts may be enriched
by a foreground structure at $z\sim 0.3$ (Ellis \etal \markcite{r14} 1985).
As can be seen in Figure 8, while the three large galaxies cause the
smoothed central light peak to be elliptical with the major axis
running north-east south-west, the fainter galaxies in the core are
distributed circularly about the BCG, with a slight over-density to the west
of the BCG.  The centroids of both the galaxy counts and galaxy
luminosities are slightly south-west of the location of the BCG.

The mass distribution created with the KS93 algorithm is shown in Figure
8.  The central peak of the mass distribution is roughly the same size
and is in the same location as both the galaxy count and luminosity peaks.
There are four ``arms'' of matter extending radially from the central peak,
and three of the arms have analogs in the galaxy count and luminosity maps.
The strength of the signal in these arms, however, is barely above the level of
the noise objects generated in the simulated fields shown in Figure 7, and
thus it is unclear how well they might indicate cluster substructure.
The shape of the central peak, however, is very similar to that of the
ROSAT HRI observations (Neumann \& B\"ohringer \markcite{r35} 1997), which is
overlayed in contours on the mass reconstruction in Figure 8.  The
offset between the peaks in the X-ray luminosity and the weak lensing
massmap is not significant and is presumably caused by the intrinsic
ellipticity distribution of the background galaxies.  Similar sized
offsets have been seen in the simulations between the reconstructed
mass peak and the true mass peak.

Smail \etal \markcite{r40} (1995) also performed a weak lensing
analysis on this 
cluster.  Their aperture densitometry profile and mass at 300 $h^{-1}$
kpc agree within errors with ours.  There is a difference in the shape
of the central peak in the mass-maps, but that can be attributed to the
difference in the background galaxy populations used to perform the
lensing analysis.  The difference between the two mass-maps is similar
to what is seen in weak lensing reconstruction simulations of the same
lensing potential using two different background galaxy populations.

\subsection{MS $0451.6-0305$}

MS $0451.6-0305$ is the most X-ray luminous cluster in the EMSS (GL).
A composite three-color image of the cluster is given in Figure 13.

The core of the cluster is easy to recognize as a large bar of galaxies
with a north-west south-east orientation.
The brightest cluster galaxy (BCG) is in the middle of the bar, but is
the second brightest galaxy in that area due to a foreground galaxy
lying just south of it.  In both the galaxy count and luminosity maps
(Figure 8)
the bar-like structure of the core is clearly visible.  The centroid of this
bar in the galaxy count map is consistent with the location of the BCG, while
the centroid of the luminosity map is slightly to the south-east of the BCG
due to the galaxies on that side generally being somewhat more
luminous than those to the north-west of the BCG.  Both maps show that
outside the core, the majority
of the cluster galaxies are also located either to the south-east or the
north-west of the BCG.

The peak of the mass is centered on the BCG, but the broad bar-like
structure evident in the galaxy distribution has a much smaller
spatial extent in the mass reconstruction.  A ROSAT HRI X-ray image of
the cluster (Donahue \markcite{r10} 1996) is overlayed in contours on the mass
reconstruction.  It also shows the bar-like structure of the core but
with a much smaller extent than that of the galaxy distribution.
A moderately large
northern extension from the central peak present in the mass reconstruction
is consistent with a (much weaker) structure seen in the cluster galaxy 
distribution.  Further, most everywhere one can find a maxima in the galaxy
distribution, a corresponding mass signal can be found, although most of these
are just barely above the level of the noise seen in the simulations.

\subsection{MS $2053.7-0449$}

MS $2053.7-0449$ has the lowest X-ray luminosity of the $z>0.5$ clusters
detected in the EMSS.  A composite three-color image of the cluster is given
in Figure 14.  MS $2053.7-0449$ is much harder to find optically than the
rest of the clusters in the catalog given that it is at a lower
galactic latitude.   
A group of moderately bright stars ($15\lesssim R\lesssim 17$)
is projected on, and partly obscures, the southern part of the cluster.
The brightest cluster galaxy (BCG) is located just above a triangle formed
by pairs of stars.  The cluster core is plainly evident in the
smoothed luminosity image and the centroid of this core is located about
6$\arcsec $ west of the
location of the BCG.  In the galaxy count image, however, the core is
extended in a bar running roughly north-south, and there is a second bar
of similar size and density located to the south of the core of the cluster.
As a complete redshift catalog has not been compiled for this cluster,
it is uncertain as to whether the southern bar structure and a second
over-density of galaxies located north-east of the core are associated with
the cluster.  

The centroid of the central peak is consistent with the location of the
BCG.  Unlike the other $z\sim 0.55$ clusters, the central peak is not a simple
elliptical, but is shaped similar to the letter ``c''.  While the simulations
shown earlier nearly always resulted in the detection of a compact core,
they were done with a lensing mass a factor of 3-4 higher than the apparent
mass of MS $2053.7-0449$ based on the X-ray luminosity.  
The smaller central potential results in a greater distortion
to the central peak, and in roughly $15\% $ of the simulations with a mass
similar to MS $2053.7-0449$'s the central peak was distorted enough that it could
no longer be considered to have an elliptical shape.  There is a mass
detection in the
region where the southern bar of galaxies is seen in the galaxy counts.

\subsection{MS $1054.4-0321$}

MS $1054.4-0321$ is the highest redshift cluster in the EMSS sample, and until
recently it was the highest redshift cluster with a detected X-ray flux
(Luppino and Gioia \markcite{r30} 1995).  
This cluster was previously analyzed in LK
using many of the same techniques as here, although
the data used were not as deep.  Thus we should have the same results
as those in LK with allowance for errors caused by a different
selection of background galaxies used in the analysis.  This provides
a good check on the removal of the distortions introduced by the Keck
focal plane.

The three color image of the cluster is given in Figure 14.
The cluster is very easy to recognize in the image as a broad swath of
galaxies running mostly east-west.  The brightest cluster galaxy (BCG)
is in the middle of the swath.   Both the color selected galaxy
luminosity and number distribution show the cluster to be long
and extended, looking similar to a short filament.  As one would expect, the
luminosity map is more sharply peaked in the core than the galaxy distribution
map, indicating that the galaxies near the core are brighter, and therefore
bigger, than the galaxies further from the center of mass for the cluster.
The centroids of both the galaxy counts and luminosities are consistent with
the position of the BCG.  When one uses a smaller smoothing scale than
that in Figure 9, the
filamentary structure of the core can be broken into three separate
peaks.  One of these peaks, the largest in both galaxy number counts
and luminosity, is centered on the BCG, and the other two are located
to the west and north-east of the BCG.

The shape of the central mass peak is extremely similar to the shapes
seen in the galaxy count and luminosity maps.  
As with the cluster galaxy number
count and luminosity images, if the mass reconstruction is smoothed on
a smaller scale than that shown in Figure 9, the mass peak becomes three
different peaks with the central peak being the largest (most
massive).  No other structures in the mass 
reconstruction are above the level of the noise seen in the simulations.

The ROSAT HRI image (Donahue \etal \markcite{r11} 1998) of the cluster
also agrees in both position and angle with the galaxy count,
luminosity, and weak lensing mass maps for the cluster with the
exception that it does not have the small peak north-east of the
cluster core and has a maximum in the western, instead of the central,
peak.  A second ROSAT image (Neumann \etal \markcite{r55} 2000) has
the western peak much smaller which suggests that this might be a
variable source or noise.  The small southern 
extension from the cluster core in the X-ray map is not seen in either
the weak lensing or color-selected galaxy maps, but could be caused
by a foreground source, possibly the star located south-west of the
cluster core (Neumann private communication).
The fact that three different techniques
of tracing mass all show the same non-circular distribution simply provides
more compelling evidence that MS $1054.4-0321$ does not have a spherical
mass distribution, and may
indicate that it is not yet virialized and is just in process of forming.

The weak lensing data presented here agrees very well with that of LK.  
The shape of the central mass peak in LK is similar to that seen above,
although with more noise due to the decreased number of
background galaxies available for the analysis.
The radial profile also has good agreement, within the errorbars, although,
as expected, the Keck data shows a slightly higher $\bar{\kappa }$, which
is indicative of the median redshift for the background galaxies being
somewhat higher than the background galaxies used in LK.

This cluster has also been recently studied using an HST mosaic image
by Hoekstra \etal \markcite{r70} (2000).  The smaller PSF in the HST
images resulted in their being able to obtain a number density of
faint galaxies roughly twice what was detected in the Keck image.
This allowed them to detect the three components of the cluster
core at a higher significance.  The weak lensing shear and aperture
densitometry profile from the HST data agree within errors with that
presented here.

\subsection{MS $1137.5+6625$ and RXJ $1716.6+6708$}

Our data on MS $1137.5+6625$ and RXJ $1716.6+6708$
were presented in an earlier paper (Clowe \etal \markcite{r07} 1998).
In the aperture densitometry profiles published in that paper,
however, we had not accounted for the breakdown of the weak lensing
approximation, and thus the given profiles are too compact.  After
applying the breakdown of the weak lensing approximation to the best
fit models for MS $1137.5+6625$, we find that it can be fit well with an
isothermal sphere, and thus we withdraw our 
assertion that the weak lensing results suggest that the cluster is
a filamentary structure extending along the line of sight.

\section{Discussion}

In the previous section we have shown that we were able to measure a
weak lensing signal from all six of the clusters in our sample.
An absolute mass measurement for each cluster cannot be currently 
obtained from this data due to the unknown redshift distribution of
the background galaxies being lensed and the small field size.  
If one assumes, however, that
the background galaxy population in each of the images has the same
redshift distribution (ie: no large overdensities of objects at a
given redshift, etc.) and the best fit profiles accurately determine
the mass density at the edges of the field, then we can compare
properties of the clusters amongst themselves.

One such comparison is that of the quality of fit of an
isothermal sphere to the shear profile of a cluster.  To do this we have
assumed that the NFW profile given in the previous section for each
cluster provides the ``best'' fit to the data,
and therefore its $\chi ^2$ represents the noise inherent in the data.
We can then perform an F-test (Bevington \&\ Robinson \markcite{r52}
1992) which 
calculates the ratio of the reduced $\chi ^2$ of the isothermal sphere
fit to that of the NFW profile fit.  Given that we can bin the data to
have the same number of bins for each cluster, a simple comparison of
the resulting ratios indicates the quality of the isothermal sphere
fit, a lower ratio meaning higher quality.  The results, given in
Table 2, show that the most massive clusters, as measured by X-ray
temperature and luminosity, are less well fit by an isothermal sphere
than are the lower mass clusters.  Based on the F-test, MS $0451.6-0305$ and
MS $1054.4-0321$ can be excluded from being fit by an isothermal sphere at the
2 and 3 $\sigma $ level respectively.  
However, because there are a multitude of
three-dimensional density profiles which have a one-dimensional
radial profile which falls as $r^{-1}$ (an isothermal sphere and a
thin rod as examples), one cannot use this to conclude that the other
clusters are isothermal spheres.  It should also be noted that the
radii over which the shears were measured are those in which an NFW
profile can greatly resemble an isothermal sphere.  If the
measurements could be extended to either larger or smaller radii, then
one could more easily distinguish between the models.  

The significances were checked
with Monte-Carlo simulations in which the background galaxies in the images
were randomly rotated while preserving their total ellipticity and
position and then sheared by the best fit NFW profile and, separately,
by the best fit isothermal sphere.  The simulations were then fit by
both NFW profiles and isothermal spheres.  For the isothermal sphere
lenses, NFW profiles provided a better fit, as measured by the reduced
$\chi ^2$s, roughly half of the time, and the F-test significances
agreed well with the percentage of simulations which exceeded the
significance.  Similarly, the percentage of simulations with shearing
by the NFW profile which had isothermal spheres providing as good or better
fits than a NFW profile was in agreement with the significances given
in Table 2.

While the above tests were
done assuming that background galaxies lie in a sheet at $z=1.5$, this
test is relatively insensitive to a change in the redshift
distribution of the background galaxies.  Changing the background
galaxy redshift distribution merely scales all the data points, and
their errors, by the
same amount, which would result in a change in the parameters of the
best fitting profiles but not in the $\chi ^2$ itself.  
Changing the inner radius cut-off of the fits, however, would have a
significant impact on the $\chi^2$ values.  In particular, if the
minimum radius were set to $\approx 300$ kpc instead of the $\approx
100$ kpc used for the above fits, then for all the clusters the
quality of fit for the best isothermal sphere model would be
indistinguishable from the best NFW profile.

In conclusion, we have detected a weak lensing signal from six
high-redshift clusters of galaxies.  We determine that the two most
massive of these clusters, based on X-ray temperature and luminosity,
are poorly fit by an isothermal sphere.  Two of the three $z\sim 0.8$
clusters have secondary mass peaks in the cluster.  One of the
$z\sim 0.5$ clusters, MS$2053.7-0449$, may have a secondary mass peak, but
we do not have enough redshifts in the system to know if the galaxies
associated with the mass peak are at the same redshift as the cluster.
We find that over-densities in color-selected cluster
galaxies nearly always correspond to over-densities in the mass
reconstructions.  It is uncertain if any of the mass over-densities
which do not have a corresponding galaxy over-density are significant,
given that they are typically near the level of noise seen in
simulations and could be caused by structures at a different redshift
whose galaxies would not appear in the color-selected galaxy catalog.
Based on these results, we caution that any attempt
to compare these clusters to those at lower redshift must take into
account that at least half of these clusters have not appeared to have
fully collapsed.

\acknowledgements
We thank Gillian Wilson, Lev Koffman, Len Cowie, Dave Sanders, 
John Learned, and Peter Schneider for their help and advice.  We also 
wish to thank Pat Henry, Harald Ebeling, Chris Mullis, Megan Donahue,
and Doris Neumann for sharing 
their X-ray data with us before publication.  This work was supported
by NSF Grants AST-9529274 and AST-9500515, Nasa Grant NAG5-2594,
ASI-CNR, and the ``Sonderforschungsbereich 375-95 f\"ur
Astro--Teil\-chen\-phy\-sik" der Deutschen
For\-schungs\-ge\-mein\-schaft.

\newpage
\begin{figure}[!hp]
\vspace{6in}
\includegraphics{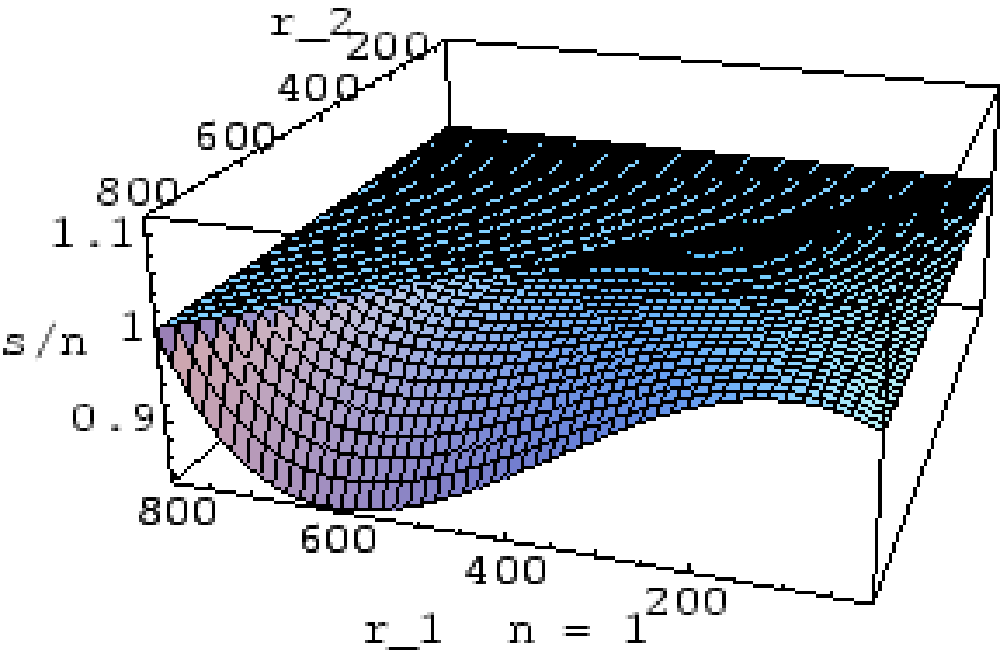}
\includegraphics{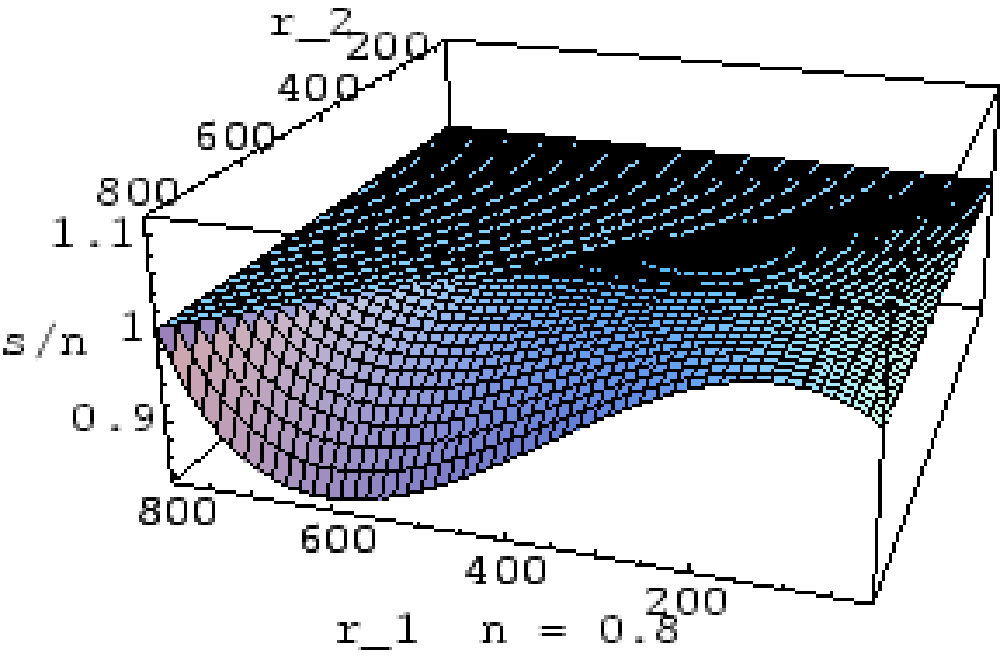}
\includegraphics{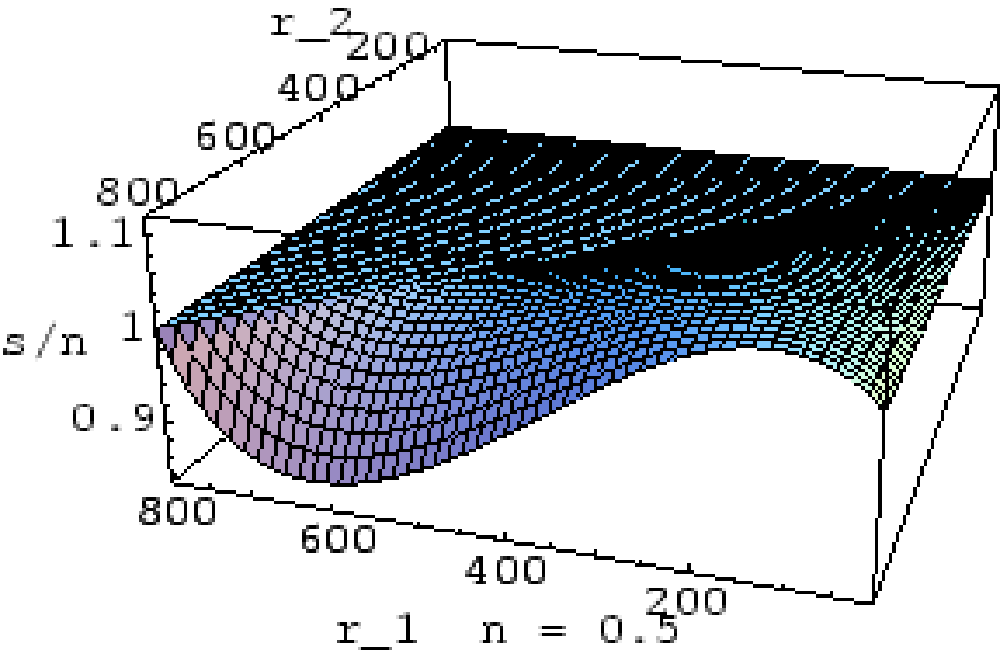}
\includegraphics{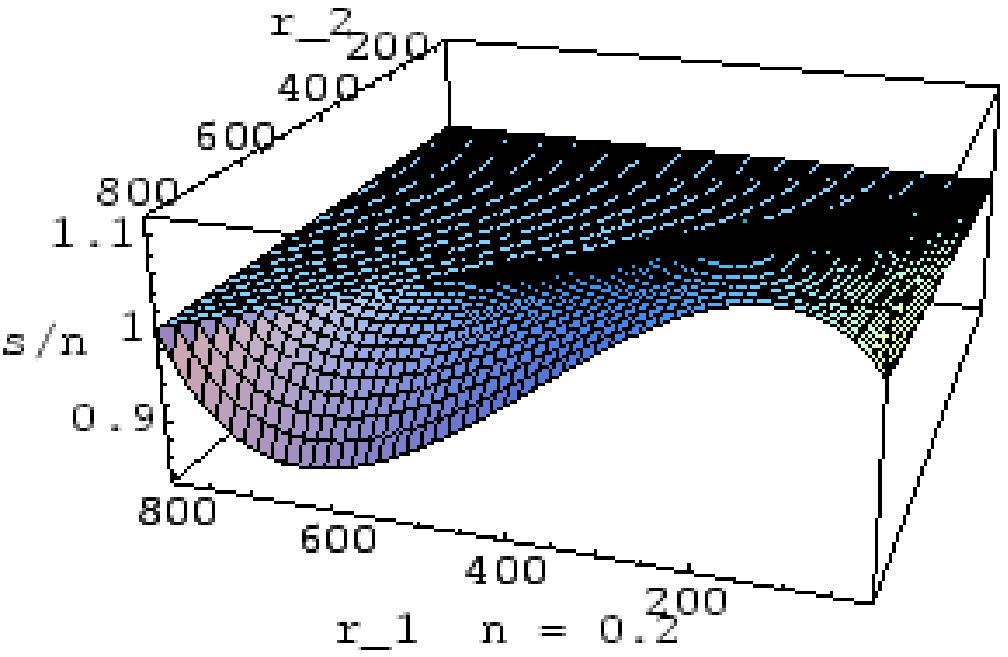}
\figcaption{Signal-to-noise ratio of the aperture densitometry
statistics.  This plot shows the ratio of ${\zeta _c/\sigma _{\zeta
_c}\over \zeta / \sigma{\zeta }}$ as a function of $r_1$ and $r_2$ for
four different signals with power-law surface densities $\kappa
\propto r^{-n}$. $r_1$ and $r_2$ are given as pixels, and $r_m$ is
assumed to be at 1000 pixels.  As can be seen, the $\zeta _c$
statistic tends to have a higher signal-to-noise than the $\zeta $
statistic at the inner radii but is worse at larger radii.  $\zeta _c$
is also better for mass profiles which are not as concentrated (low $n$).}
\end{figure}

\clearpage
\begin{figure}[!hp]
\vspace{6in}
\includegraphics{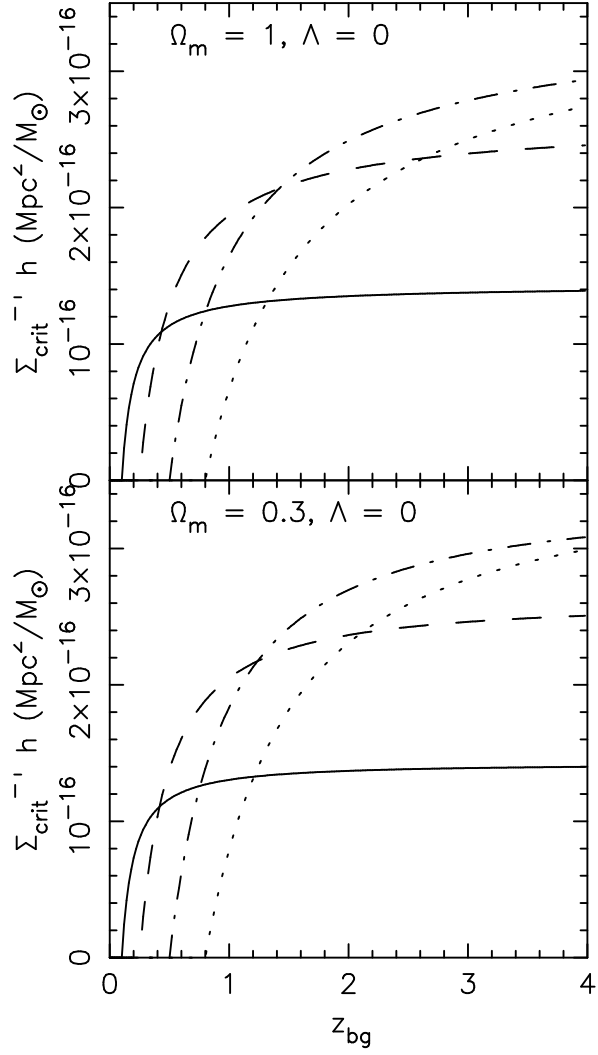}
\figcaption{Lensing strength as a function of background galaxy
redshift.  The above graphs show lensing strength ($\Sigma
_{crit}^{-1}$) as a function of background galaxy redshift for four
cluster lens redshifts ($z = 0.1, 0.25, 0.5,$ and $0.8$).  The top
graph is for a cosmology with $\Omega _m = 1$ and the bottom graph is
for a cosmology with $\Omega _m = 0.3$ (both assume $\Lambda = 0$).
The cluster lens redshifts for each line can be determined
by where the lensing strength reaches 0.}
\end{figure}

\clearpage
\begin{figure}[!hp]
\vspace{6in}
\includegraphics{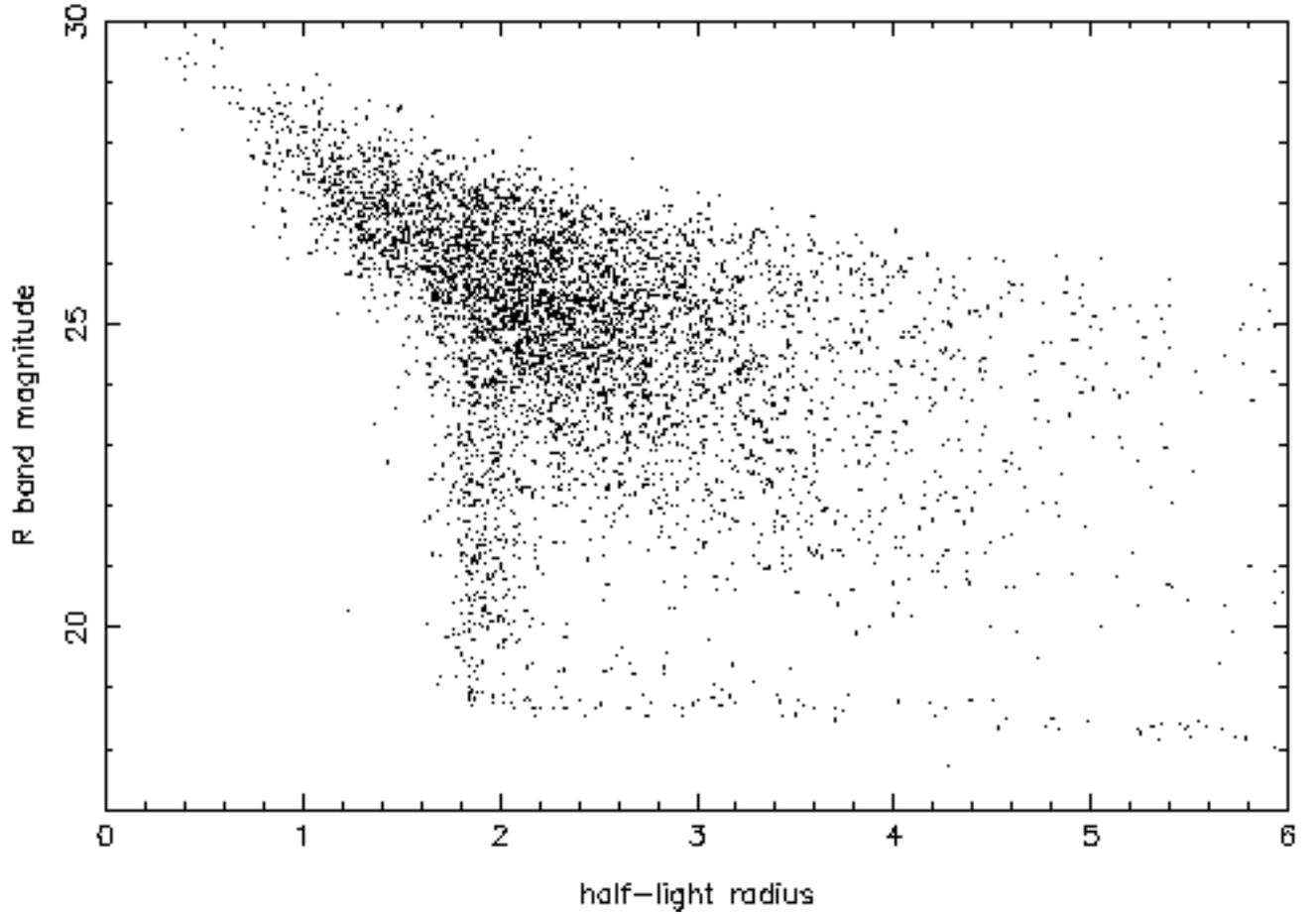}
\figcaption{A plot of $r_h$, the radius which encloses half of the
detected light from a object, versus the total magnitude of the
objects detected in the MS $2053.7-0449$ field is shown above.
Galaxies (foreground, background, and cluster) form a broad swath across the
plot while stars, all having roughly the same measured half-light radius, 
lie in a tight finger in plot.  By using this plot, moderately bright stars
can easily be separated from galaxies, although faint galaxies and stars
blend together.}
\end{figure}

\clearpage
\begin{figure}[!hp]
\vspace{6in}
\includegraphics{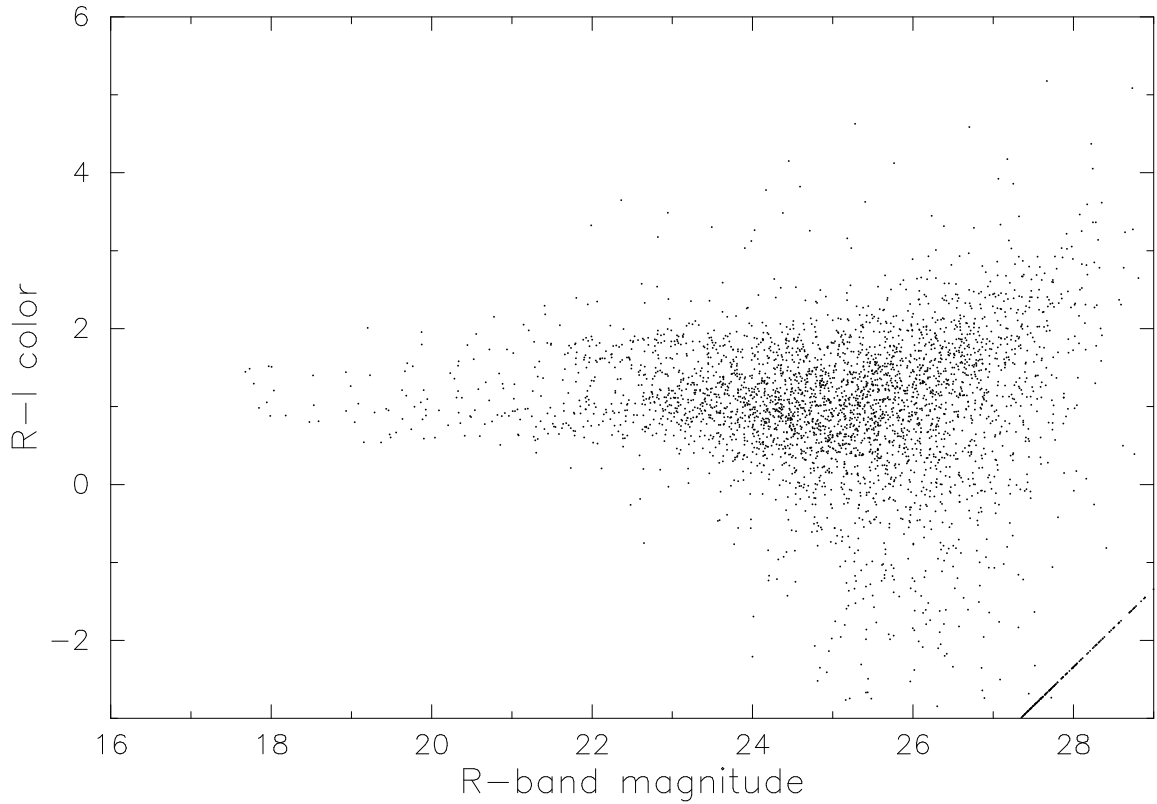}
\figcaption{A plot of the
$R$-band magnitude versus $R-I$ color for all detected objects in the 
MS $1054.4-0321$ field is shown above.  The brighter cluster galaxies can be
detected as the narrow finger in color around $R-I = 1.6$.  As can be seen,
at fainter magnitudes the galaxies split into two populations based on color.
The red population tends to blend with the fainter cluster galaxy population,
which will tend to dilute any lensing signal for those galaxies.  In order
to exclude cluster galaxies from the background galaxy sample only galaxies
with $R-I <$ 0.9 will be used in the analysis.}
\end{figure}

\clearpage
\begin{figure}[!hp]
\vspace{6in}
\includegraphics{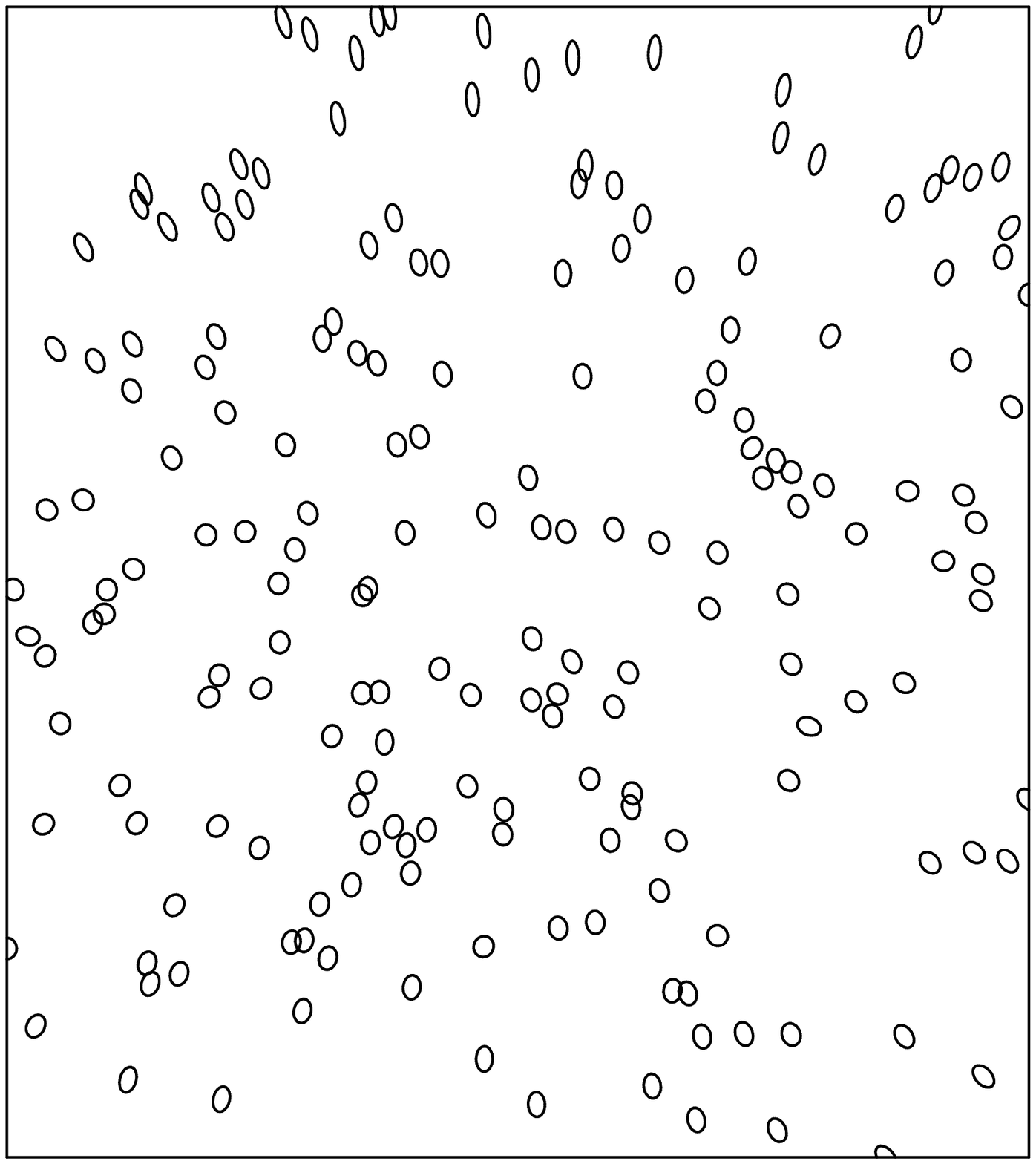}
\includegraphics{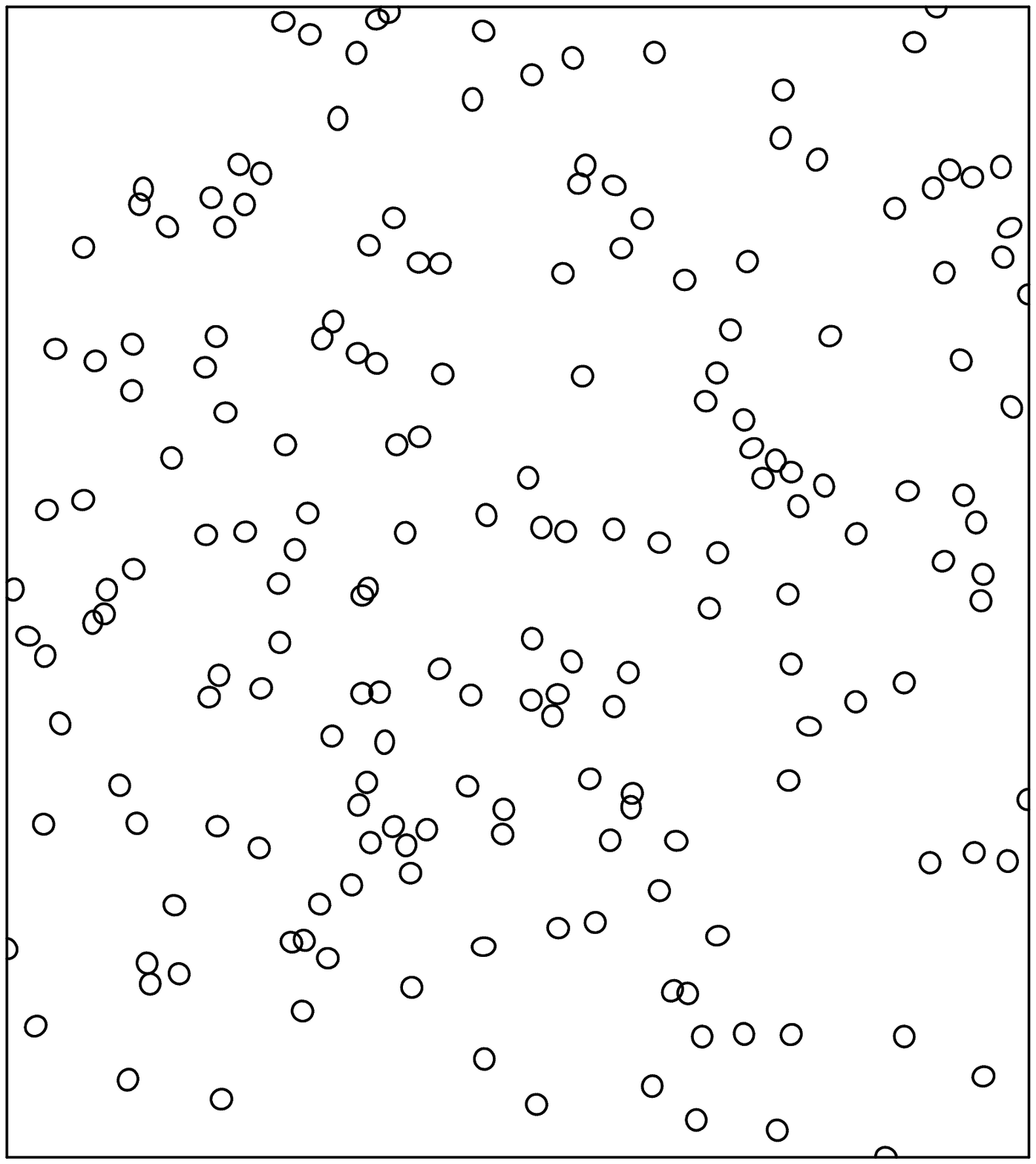}
\figcaption{The above plots show the ellipticities of the bright but
unsaturated stars in the MS $2053.7-0449$ $R$-band Keck image.  
The ellipticities have been magnified by a factor of five in order to clearly
demonstrate the effect seen.  The image on the left shows the
original ellipticities, which show the effects of the astigmatism left
in the Keck images after correction for the curvature of the focal
plane.  The image on the right shows the residual ellipticities after a
bi-cubic polynomial has been fit to and subtracted from the stellar
ellipticities.}
\end{figure}

\clearpage
\begin{figure}[!hp]
\vspace{6in}
\includegraphics{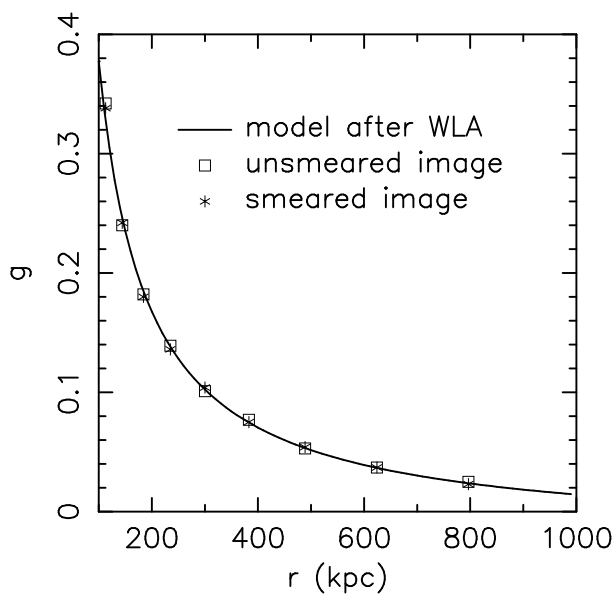}
\figcaption{The above graph shows the average mass
profile created from aperture densitometry of 20 simulations of a background 
galaxy field lensed by massive cluster.  The input mass profile
(corrected for the breakdown of the weak lensing approximation) is drawn as
a solid line.  The open squares are the aperture densitometry profile
using the sheared background galaxies without any smearing by a psf,
and thus represent the noise caused by the intrinsic ellipticity
distribution of the galaxies.  The stars are the aperture densitometry
profile using the sheared galaxies which were smeared by a psf taken
from the MS $2053.7-0449$ Keck $R$-band image, and corrected by the KSB
method described in \S $3.4$.}
\end{figure}

\clearpage
\begin{figure}[!hp]
\vspace{6.5in}
\includegraphics{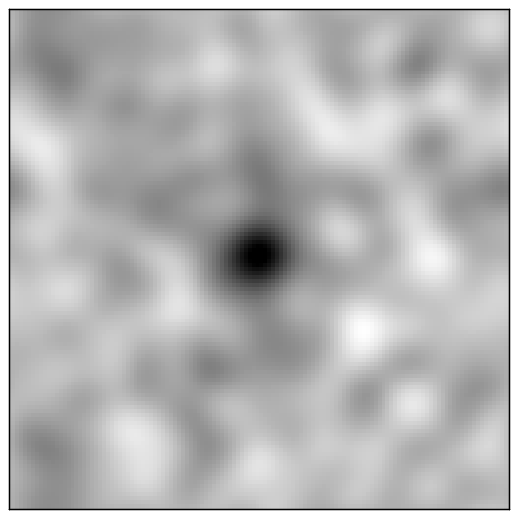}
\includegraphics{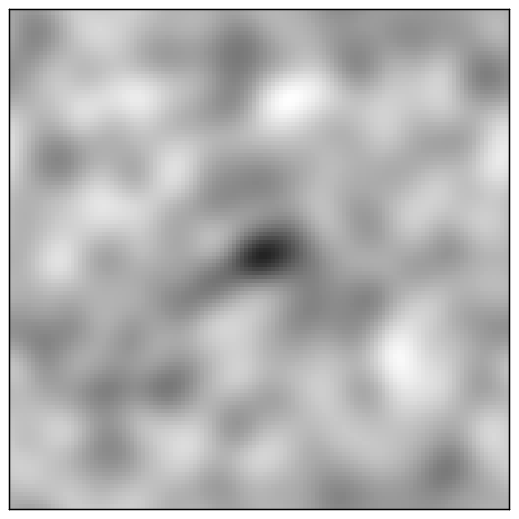}
\includegraphics{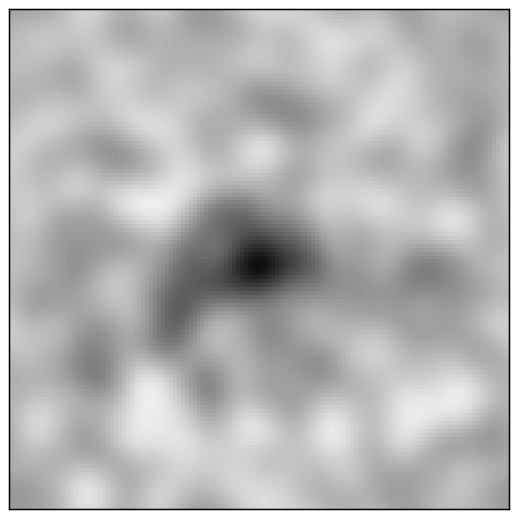}
\includegraphics{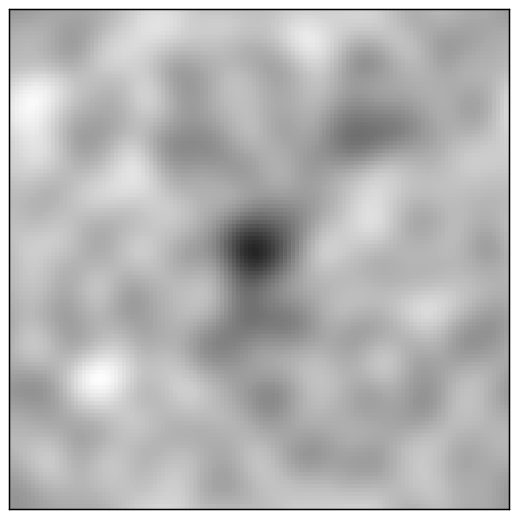}
\includegraphics{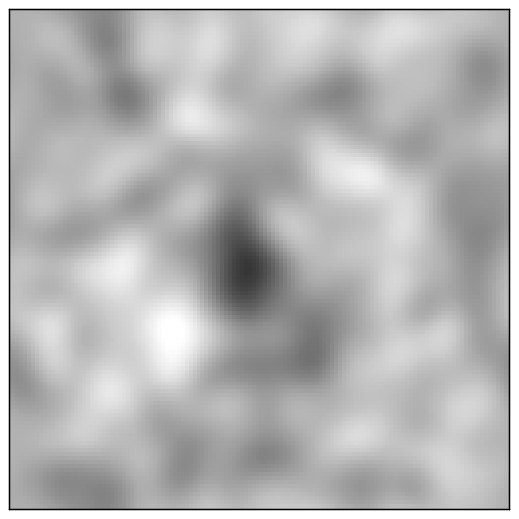}
\includegraphics{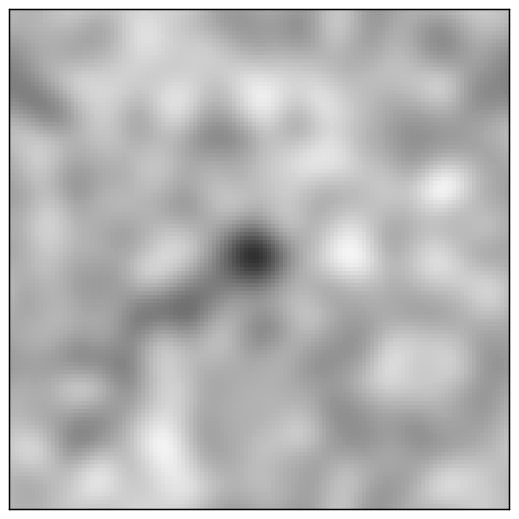}
\includegraphics{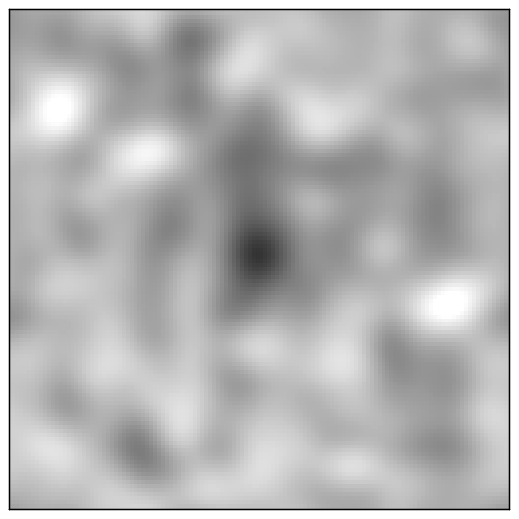}
\includegraphics{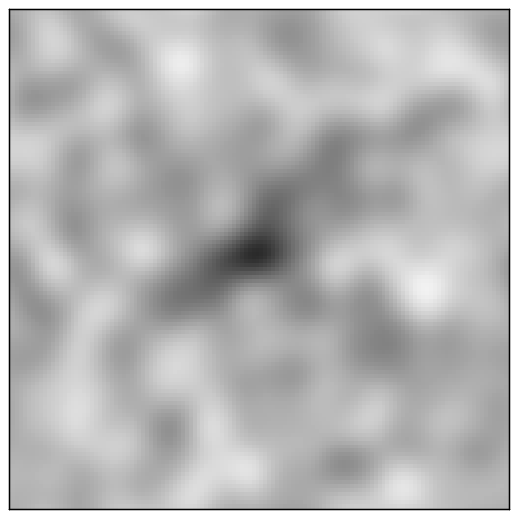}
\includegraphics{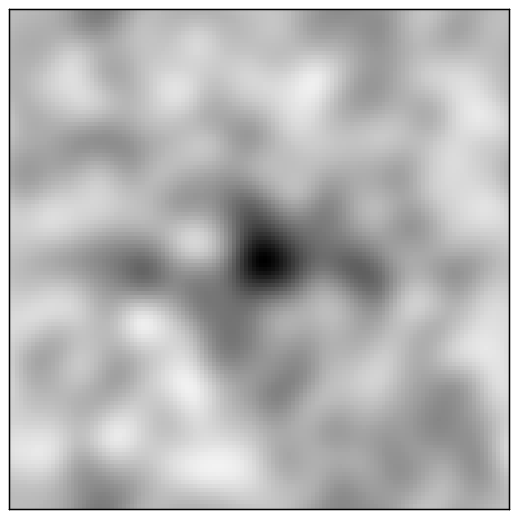}
\figcaption{Shown above are nine mass reconstructions generated with
the KS 93 algorithm for a simulated isothermal lens ($\sigma = 1000$
km/s, $z_{lens} = 0.8, z_{bg} = 1.5$).  As can be seen, the central
peak is always distinguishable above the
level of the noise, but is usually somewhat elliptical in shape, and often
displaced by a small amount from the center of the field.  All of the above
images are of the same model lens, the differences in the images are a result
of the random ellipticities of the background galaxies.}
\end{figure}

\clearpage
\begin{figure}[!hp]
\vspace{7in}
\includegraphics{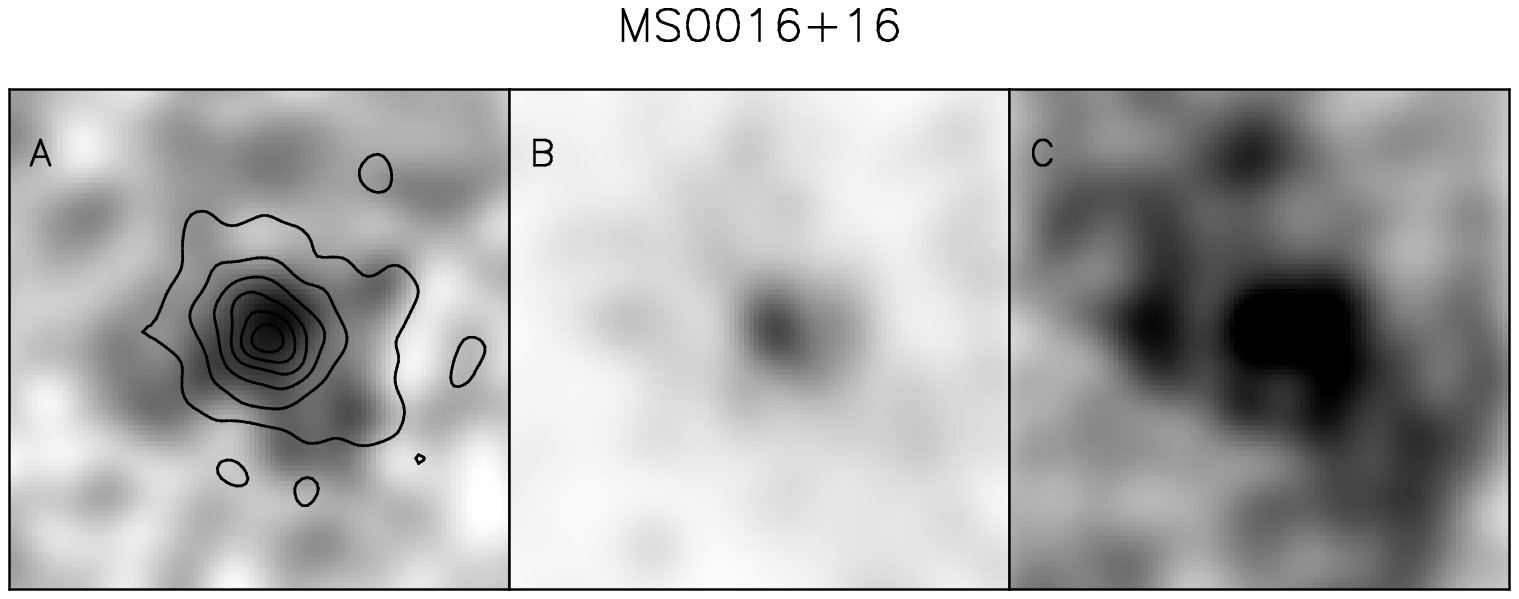}
\includegraphics{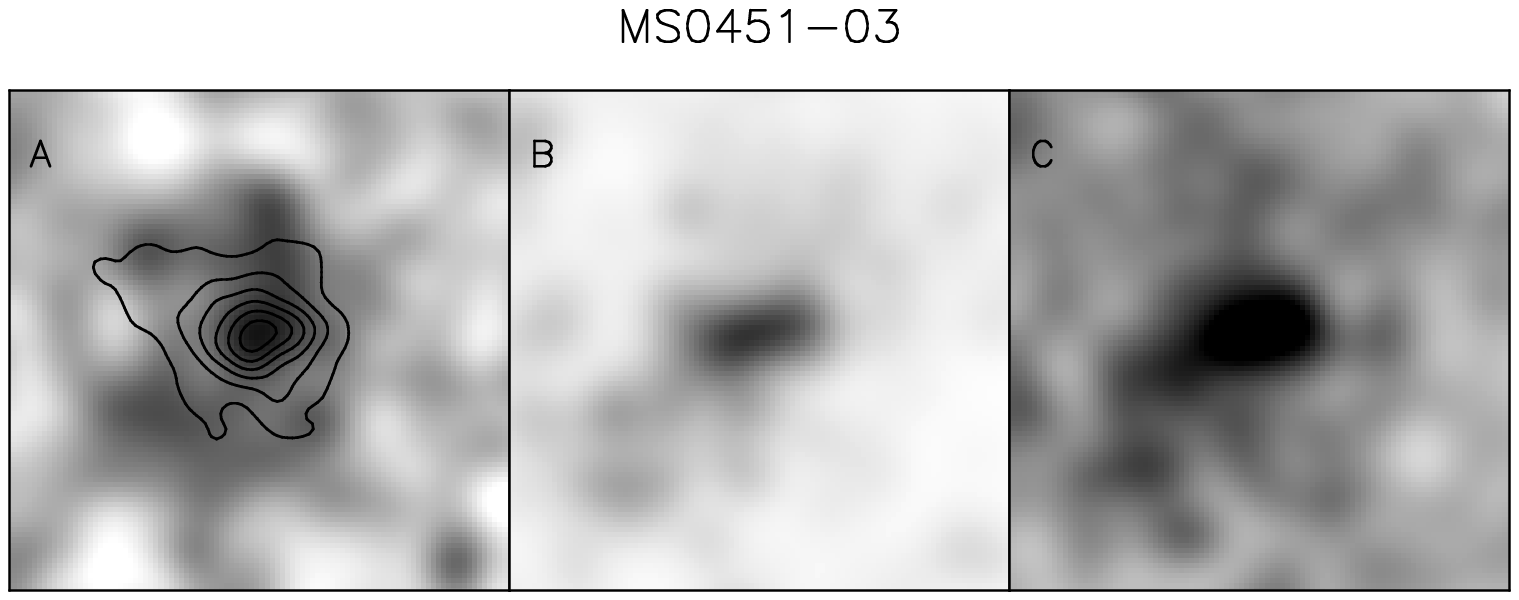}
\includegraphics{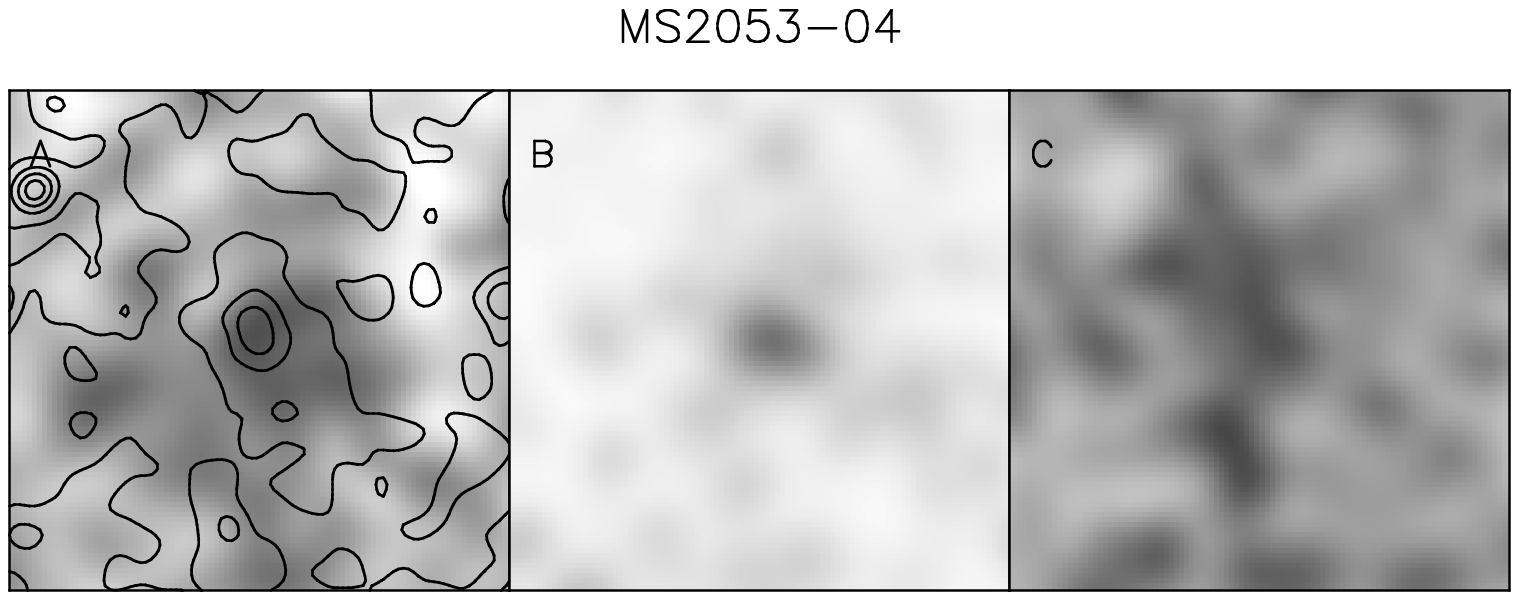}
\figcaption{Mass surface density, color-selected galaxy luminosity and
number count map of the $z\sim 0.55$ clusters are shown above.  The
mass reconstructions from the KS93 algorithm are labeled A, and the
galaxy luminosity and number count maps are labeled B and C
respectively.  The maps are all $5\farcm9$ on a side (same as the
color images in Figures 12-15) and have all been smoothed by a
17\arcsec Gaussian.  The grey-scale of the maps is the same in all
maps displaying the same quantity (all mass reconstructions, {\it
etc.}).}
\end{figure}

\clearpage
\begin{figure}[!hp]
\vspace{7in}
\includegraphics{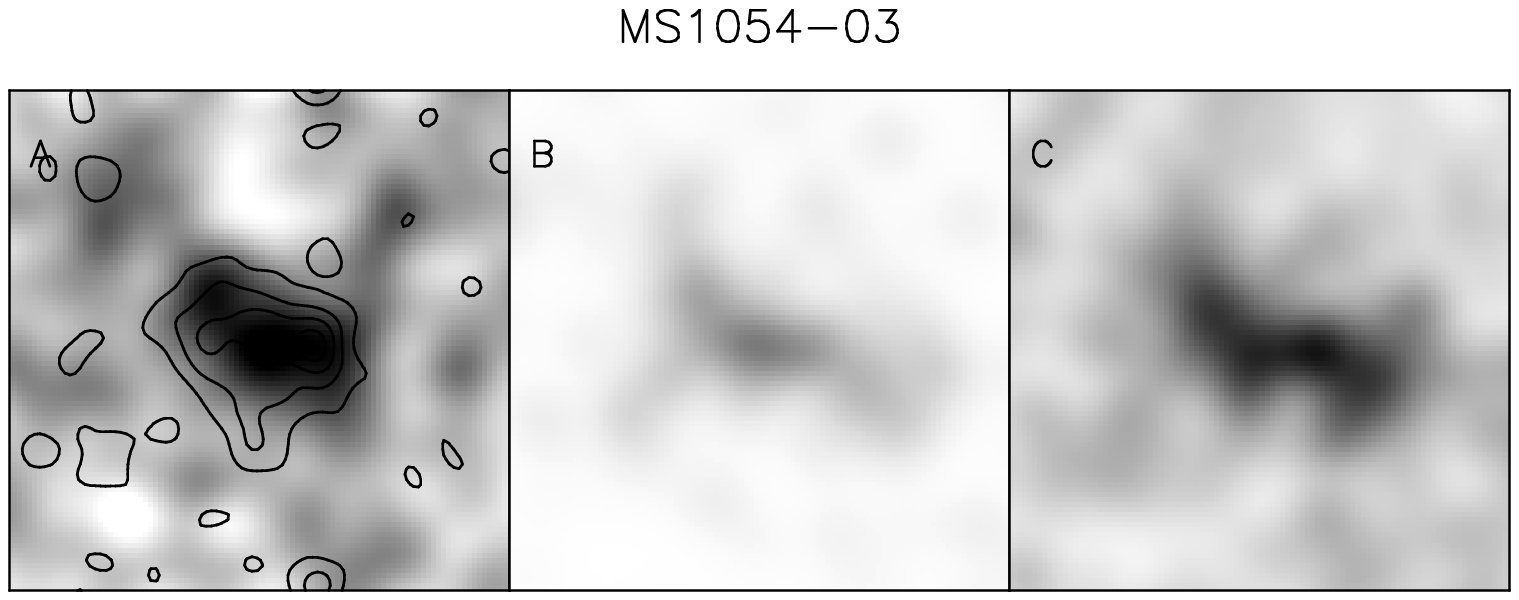}
\includegraphics{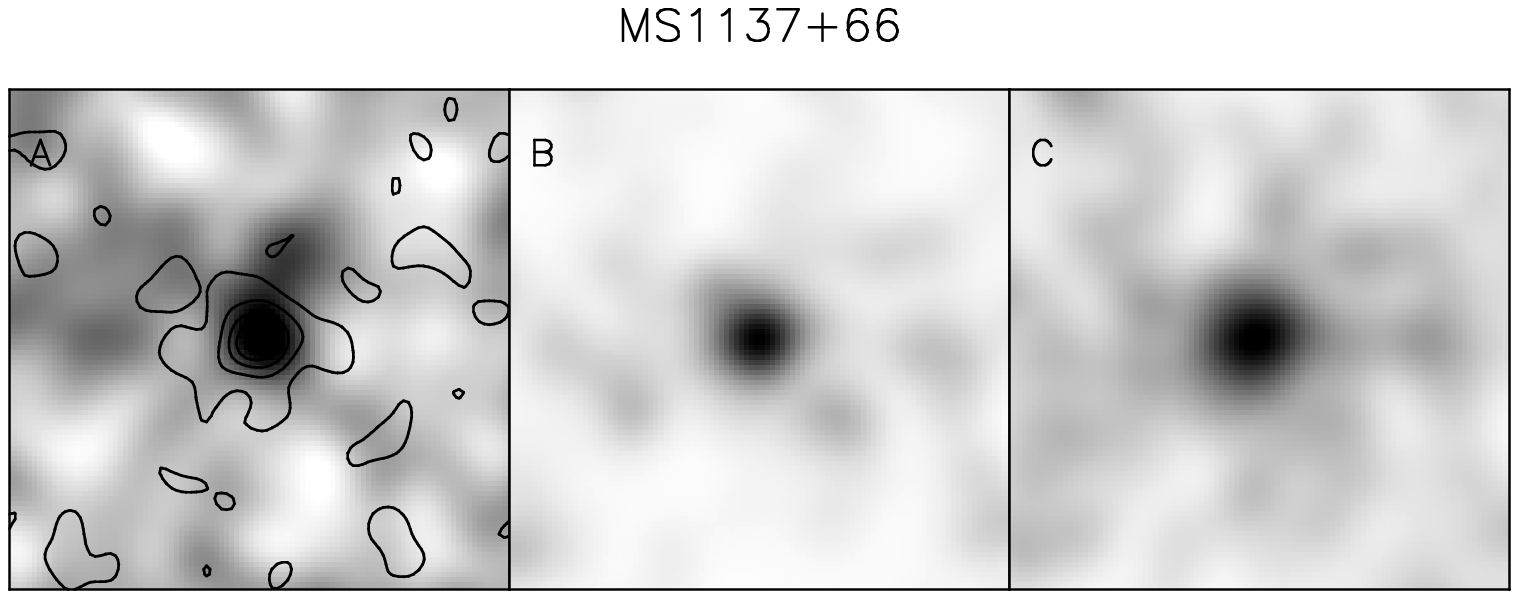}
\includegraphics{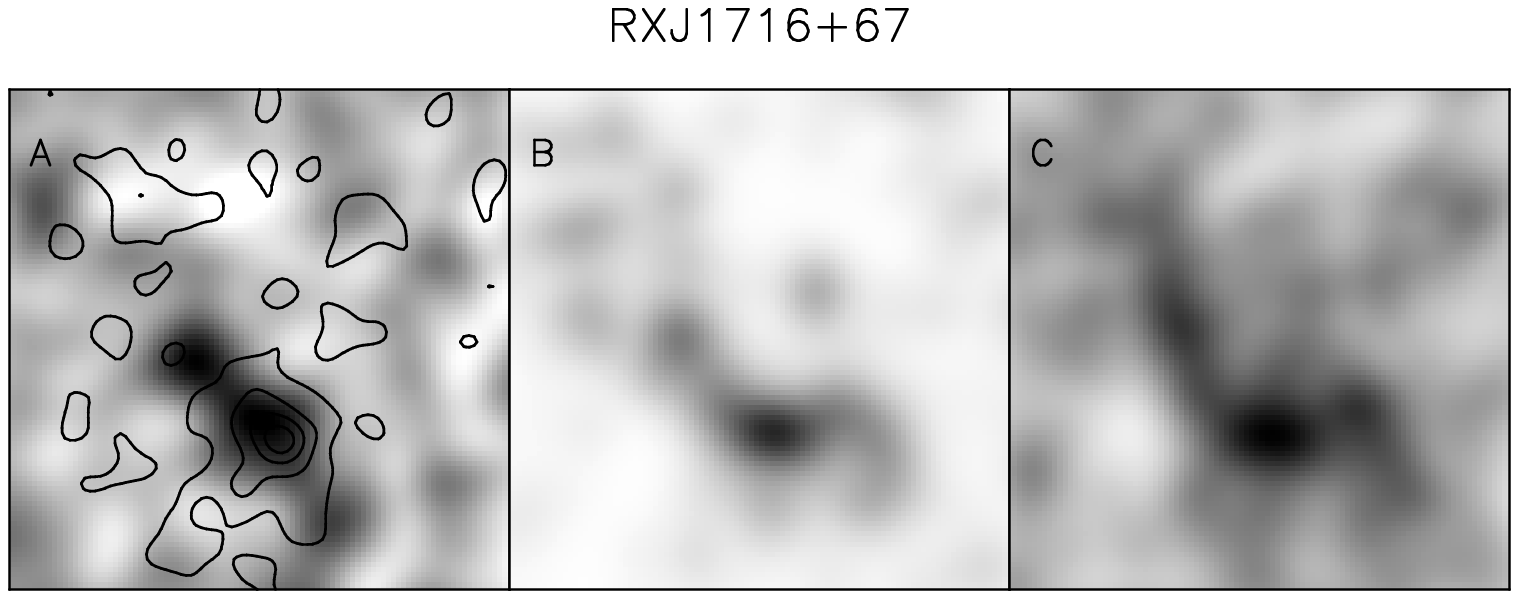}
\figcaption{Mass surface density, color-selected galaxy luminosity and
number count map of the $z\sim 0.8$ clusters are shown above.  The
mass reconstructions from the KS93 algorithm are labeled A, and the
galaxy luminosity and number count maps are labeled B and C
respectively.  The maps are all $5\farcm9$ on a side (same as the
color images in Figures 12-15) and have all been smoothed by a
17\arcsec Gaussian.  The grey-scale of the maps is the same in all
maps displaying the same quantity (all mass reconstructions, {\it
etc.}) and are the same as in Figure 8.}
\end{figure}

\clearpage
\begin{figure}[!hp]
\vspace{6in}
\includegraphics{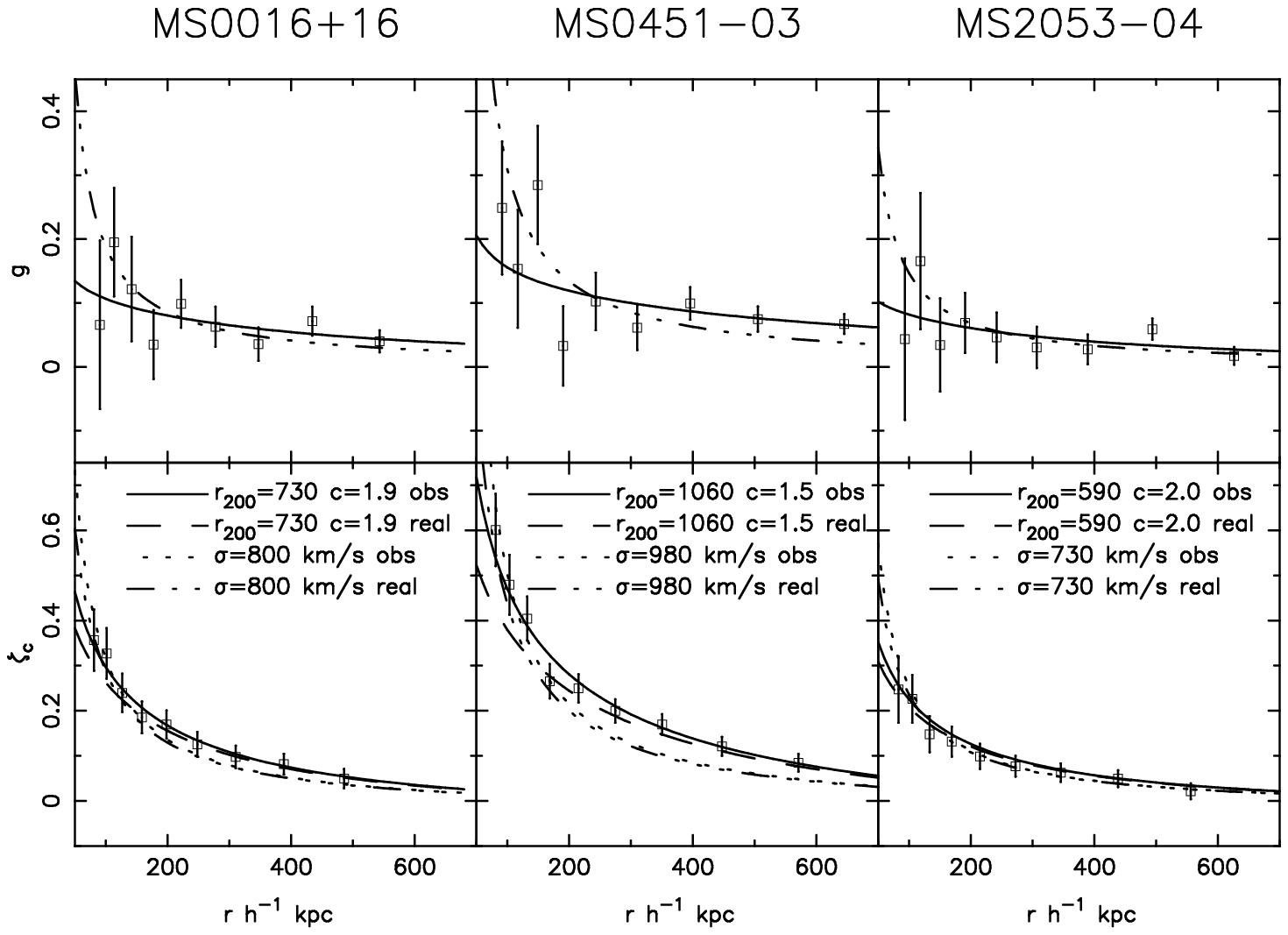}
\includegraphics{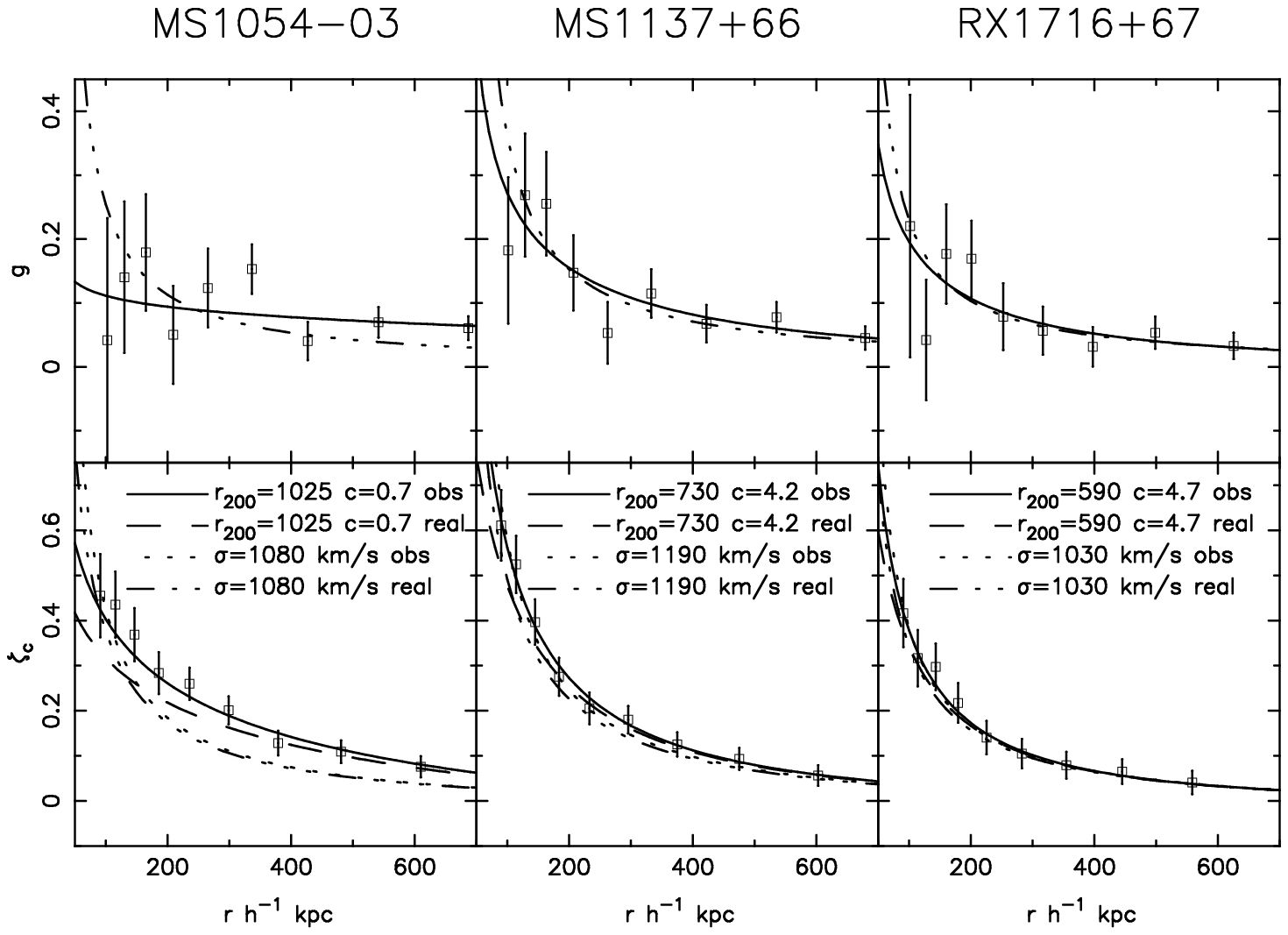}
\figcaption{The shear and aperture densitometry profiles for all of
the clusters in the sample is shown above.  The top figure in each
pair shows the detected reduced shear tangential to the vector from the BCG to
the galaxy being measured.  The galaxies have been placed in radial bin to
reduce the noise from the intrinsic ellipticity distribution of the
background galaxies.  The bottom figure shows the value of
$\zeta _c$ calculated from
the shear values above along with two models for the mass profile of
the cluster.  Best fit profiles (fitting done to the shear values) of
both an isothermal sphere and an ``universal'' CDM profile are shown.
Given are both the true profiles of the models and those which are
measured due to the breakdown of the weak lensing approximation, the
difference of which can be used to estimate the effect of the
breakdown on the mass estimates if it were to be ignored.  The
data points in the shear plot are all independent, but as $\zeta _c$ is
created by summing the shear values divided by the radius, the $\zeta
_c$ data points are not independent.} 
\end{figure}

\clearpage
\begin{figure}[!hp]
\vspace{6in}
\includegraphics{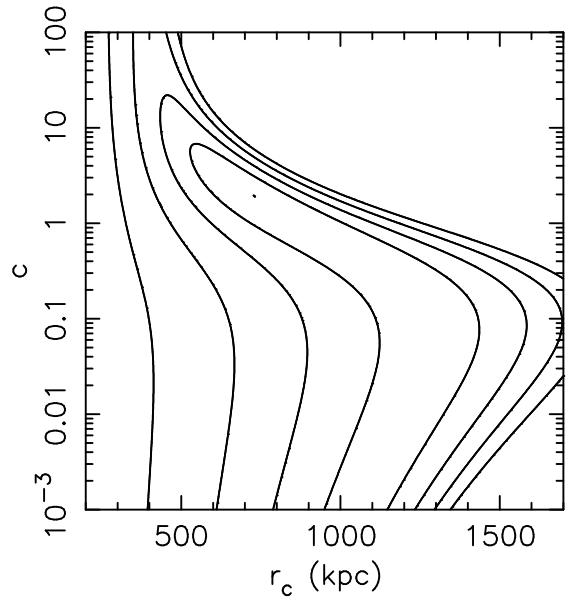}
\figcaption{$\chi ^2$ values are plotted as contours above for
``universal'' CDM profile fits to the MS $0015.9+1609$ shear profile.
Each contour represents a change of $1\sigma $ in the quality of fit.
As can be seen, the two input parameters can be traded-off to some
extent without severely effecting the quality of the fit.}
\end{figure}

\clearpage
\figcaption{$5\farcm 9\times 5\farcm 9$, 3 color image of 
MS $0015.9+1609$.  R, G, and B colors are
the 29700s $I$-band exposure from the UH88'' telescope, 4200s $R$-band 
exposure from the Keck II telescope, and 19800s $B$-band exposure from 
the UH88'' telescope respectively.  All three
colors are scaled with a $\log ^{1/2}$ stretch.}

\figcaption{$5\farcm 9\times 5\farcm 9$, 3 color image of 
MS $0451.6-0305$.  R, G, and B colors are
the 18900s $I$-band exposure from the UH88'' telescope, 6600s $R$-band 
exposure from the Keck II telescope, and 14400s $B$-band exposure from 
the UH88'' telescope respectively.  All three
colors are scaled with a $\log ^{1/2}$ stretch.}

\figcaption{$5\farcm 9\times 5\farcm 9$, 3 color image of 
MS $2053.7-0449$.  R, G, and B colors are
the 35100s $I$-band exposure from the UH88'' telescope, 4500s $R$-band 
exposure from the Keck II telescope, and 15300s $B$-band exposure from 
the UH88'' telescope respectively.  All three
colors are scaled with a $\log ^{1/2}$ stretch.}

\figcaption{$5\farcm 9\times 5\farcm 9$, 3 color image of 
MS $1054.4-0321$.  R, G, and B colors are
the 21600s $I$-band exposure from the UH88'' telescope, 6300s $R$-band 
exposure from the Keck II telescope, and 10800s $B$-band exposure from 
the UH88'' telescope respectively.  All three
colors are scaled with a $\log ^{1/2}$ stretch.}

\newpage
\begin{deluxetable}{lcccc}
\tablewidth{5.0in}
\footnotesize
\tablehead{\colhead{Cluster Name}&\colhead{Band}&\colhead{Date}&
	\colhead{$T_{obs}$}&\colhead{seeing}}
\startdata
MS$0015.9+1609$&$R$&Aug 17-18 1996&4200s&0\farcs 6\nl
         &$I$&Sep 10-12 1996&29700s&0\farcs 7\nl
	 &$B$&Sep 4-5 1997&19800s&0\farcs 7\nl
\hline MS$0451.6-0305$&$R$&Jan 10-11 1997&6600s&0\farcs 8\nl
         &$I$&Sep 10-12 1996&18900s&0\farcs 7\nl
	 &$B$&Sep 4-5 1997&14400s&1\farcs 0\nl
\hline MS$1054.4-0321$&$R$&Jan 10-11 1997&6300s&0\farcs 8\nl
	 &$I$&Jan 11-13 1994&21600s&0\farcs 9\nl
	 &$B$&Apr 27-May 1 1997&10800s&0\farcs 7\nl
\hline MS$1137.5+6625$&$R$&Jan 10-11 1997&8700s&0\farcs 8\nl
	 &$I$&Apr 6-7 1995&8400s&0\farcs 7\nl
	 &$I$&Apr 29-May 1 1997&12600s&0\farcs 6\nl
	 &$B$&Apr 29-May 1 1997&12600s&0\farcs 9\nl
\hline MS$2053.7-0449$&$R$&Aug 17-18 1996&4500s&0\farcs 6\nl
	 &$I$&Jul 20-22 1996&19800s&0\farcs 6\nl
         &$I$&Sep 10-12 1996&15300s&0\farcs 6\nl
	 &$B$&Sep 4-5 1997&15300s&0\farcs 9\nl
\hline RXJ$1716.6+6708$&$R$&Aug 17-18 1996&7500s&0\farcs 7\nl
	 &$I$&Jul 20-22 1996&26100s&0\farcs 8\nl
	 &$B$&Apr 29-May 1 1997&10800s&1\farcs 0\nl
\enddata
\end{deluxetable}

\newpage
\begin{deluxetable}{lcccccc}
\tablewidth{6.0in}
\footnotesize
\tablehead{\colhead{~}&\colhead{MS0016}&\colhead{MS0451}&\colhead{MS1054}&\colhead{MS1137}&\colhead{MS2053}&\colhead{RXJ1716}}
\startdata
redshift&0.547\tablenotemark{a}&0.550\tablenotemark{a}&0.833\tablenotemark{b}&
0.783\tablenotemark{a}&0.586\tablenotemark{a}&0.809\tablenotemark{c}\nl
$L_x^{0.3-3.5\mbox{keV}}\times 10^{44} h^{-2}$erg/s&3.67\tablenotemark{a}&
5.00\tablenotemark{a}&2.30\tablenotemark{a}&1.90\tablenotemark{a}&
1.45\tablenotemark{a}&1.50\tablenotemark{c}\nl
$T_x$ (keV)&$9.9^{+1.1}_{-1.0}$\tablenotemark{d}&
$10.9\pm 1.2$\tablenotemark{e}&$12.3^{+3.7}_{-2.1}$\tablenotemark{f}&
$5.7^{+2.1}_{-1.1}$\tablenotemark{g}&$8.1^{+3.7}_{-2.2}$\tablenotemark{h}&
$5.7^{+1.4}_{-0.6}$\tablenotemark{c}\nl
$v$ (km/s)&1234\tablenotemark{i}&1371\tablenotemark{i}&1170\tablenotemark{b}&
884\tablenotemark{g}&...&1522\tablenotemark{c}\nl
BCG $R$ mag\tablenotemark{j} ($<10h^{-1}$kpc)&19.98&19.84&21.87&21.85&
21.80&21.97\nl
$N_{gal}$ ($r\le 250h^{-1}$ kpc)&$66^{+8}_{-7}$&$47^{+8}_{-6}$&
$44^{+7}_{-6}$&$27^{+7}_{-5}$&$20^{+4}_{-5}$&$29^{+5}_{-4}$\nl
Abell Class&IV&III&III&II&I&II\nl
$n_{bg}$ (galaxies/sq arcmin)&33.9&35.4&42.2&40.4&33.4&37.0\nl
\multicolumn{7}{l}{Best fit NFW profile}\nl
$r_{200}$ (kpc)&730&1060&1025&730&590&590\nl
c&1.9&1.5&0.7&4.2&2.0&4.7\nl
$\chi^2$(dof)&18.5(19)&26.4(20)&18.7(20)&23.6(19)&19.9(20)&19.4(19)\nl
significance from $r_{200}=0$&5.3&9.4&5.7&7.6&3.8&4.5\nl
\multicolumn{7}{l}{Best fit Isothermal Sphere}\nl
$\sigma$ (km/s)&800&980&1080&1190&730&1030\nl
$\chi^2$(dof)&20.6(20)&33.9(21)&27.9(21)&23.7(20)&20.7(21)&19.7(20)\nl
significance from $\sigma =$ 0km/s&5.4&7.8&5.1&7.2&4.2&4.9\nl
\multicolumn{7}{l}{F-test for significance between NFW and IS $\chi^2$'s}\nl
(IS-NFW)/(NFW/dof)&2.16&5.68&9.84&0.08&0.80&0.29\nl
significance (\%)&85&97&99.5&--&--&--\nl
\enddata
\tablenotetext{a}{Luppino \&\ Gioia \markcite{r30} 1995}
\tablenotetext{b}{Tran \etal \markcite{r65} 1999}
\tablenotetext{c}{Gioia \etal \markcite{r19} 1999}
\tablenotetext{d}{Yamashita \markcite{r46} 1994}
\tablenotetext{e}{Donahue \markcite{r10} 1996}
\tablenotetext{f}{Donahue \etal \markcite{r11} 1998}
\tablenotetext{g}{Donahue \etal \markcite{r61} 1999}
\tablenotetext{h}{Henry \markcite{r21} 2000}
\tablenotetext{i}{$\omega_m=0.1,\Lambda =0$, Carlberg \etal
\markcite{r06} 1999}
\tablenotetext{j}{$\pm 0.05$ magnitude error for all}
\end{deluxetable}
\end{document}